\documentclass[]{pasj02} 
\usepackage[switch,mathlines]{lineno} 

\usepackage{xcolor}
\jyear{2025}
\usepackage{url}

\usepackage{comment}

\newcommand{\iac}{Instituto de Astrof\'\i sica de Canarias (IAC), 38205 La Laguna, Tenerife, Spain}
\newcommand{\komaba}{Department of Multi-Disciplinary Sciences, The University of Tokyo, 3-8-1 Komaba, Meguro, Tokyo 153-8902, Japan}
\newcommand{\komabains}{Komaba Institute for Science, The University of Tokyo, 3-8-1 Komaba, Meguro, Tokyo 153-8902, Japan}
\newcommand{\abc}{Astrobiology Center, 2-21-1 Osawa, Mitaka, Tokyo 181-8588, Japan}
\newcommand{\naoj}{National Astronomical Observatory of Japan, 2-21-1 Osawa, Mitaka, Tokyo 181-8588, Japan}
\newcommand{\astron}{Department of Astronomy, The University of Tokyo, 7-3-1 Hongo, Bunkyo, Tokyo 113-0033, Japan}
\newcommand{\sokendai}{Department of Astronomical Science, The Graduate University for Advanced Studies (SOKENDAI), 2-21-1 Osawa, Mitaka, Tokyo 181-8588, Japan}
\newcommand{\ep}{Department of Earth and Planetary Science, The University of Tokyo, 7-3-1 Hongo, Bunkyo, Tokyo 113-0033, Japan}
\newcommand{\utops}{UTokyo Organization for Planetary and Space Science,The University of Tokyo, 7-3-1 Hongo, Bunkyo, Tokyo 113-0033, Japan}
\newcommand{\ritsumei}{Department of Physical Sciences, Ritsumeikan University, 
1-1-1 Noji-higashi, Kusatsu, Shiga 525-8577, Japan}
\newcommand{\subaru}{Subaru Telescope, National Astronomical Observatory of Japan, 650 N. Aohoku Place, Hilo, HI 96720, USA}

\newcommand{\elsi}{Earth-Life Science Institute (ELSI), Institute of Science Tokyo, 2-12-1 Ookayama, Meguro, Tokyo 152-8551, Japan}

\newcommand{\isas}{Institute of Space and Astronautical Science, Japan Aerospace Exploration Agency, 3-1-1 Yoshinodai, Chuo, Sagamihara, Kanagawa, 252-5210, Japan}

\newcommand{\lagna}{Departamento de Astrof\'{i}sica, Universidad de La Laguna (ULL), E-38206 La Laguna, Tenerife, Spain}

\newcommand{\gron}{Kapteyn Astronomical Institute, University of Groningen, P.O. Box 800, 9700 AV Groningen, Netherlands}

\newcommand{\sun}{\solar}

\begin{document} 
\title{The mass of TOI-1883~b: A low density super-Neptune in the ridge regime transiting an early-M dwarf}

\author{
 Izuru \textsc{Fukuda},\altaffilmark{1}\altemailmark\orcid{0000-0002-9436-2891} 
 Norio \textsc{Narita},\altaffilmark{2,3,4}\orcid{0000-0001-8511-2981}
 Akihiko \textsc{Fukui},\altaffilmark{2,3}\orcid{0000-0002-4909-5763}
 Teruyuki \textsc{Hirano},\altaffilmark{4,5,6}\orcid{0000-0003-3618-7535}
 Masayuki \textsc{Kuzuhara},\altaffilmark{4,5,6}\orcid{0000-0002-4677-9182}
 Hiroyuki \textsc{Kurokawa},\altaffilmark{1,7}\orcid{0000-0003-1965-1586}
 Kai \textsc{Ikuta},\altaffilmark{8}\orcid{0000-0002-5978-057X}
 Jerome P. \textsc{de Leon},\altaffilmark{2}\orcid{0000-0002-6424-3410}
 Takuya \textsc{Takarada},\altaffilmark{4,5}\orcid{0009-0006-9082-9171}
 Hiroki \textsc{Harakawa},\altaffilmark{9}\orcid{0000-0002-7972-0216}
  Hiroyuki Tako \textsc{Ishikawa},\altaffilmark{10}\orcid{0000-0001-6309-4380}
  Yasunori \textsc{Hori},\altaffilmark{11,4}\orcid{0000-0003-4676-0251}
  Tadahiro \textsc{Kimura},\altaffilmark{12,13}\orcid{0000-0001-8477-2523}
  Takanori \textsc{Kodama},\altaffilmark{14}\orcid{0000-0001-9032-5826}
 Masahiro \textsc{Ikoma},\altaffilmark{4,5,7}\orcid{0000-0002-5658-5971} 
 Akitoshi \textsc{Ueda},\altaffilmark{4,5,6}
 Aoi \textsc{Takahashi},\altaffilmark{15} \orcid{0000-0003-3881-3202}
 Enric \textsc{Palle},\altaffilmark{3,16}\orcid{0000-0003-0987-1593}
 Felipe \textsc{Murgas},\altaffilmark{3,16}\orcid{0000-0001-9087-1245}
 Gaia \textsc{Lacedelli},\altaffilmark{3,16}\orcid{0000-0002-4197-7374}
 Hannu \textsc{Parviainen},\altaffilmark{16,3}\orcid{0000-0001-5519-1391}
 John H. \textsc{Livingston},\altaffilmark{4,5,6}\orcid{0000-0002-4881-3620}
 Jun \textsc{Nishikawa},\altaffilmark{5,6,4}\orcid{0000-0001-9326-8134}
 Keisuke \textsc{Isogai},\altaffilmark{1,17}\orcid{0000-0002-6480-3799}
 Kiyoe \textsc{Kawauchi},\altaffilmark{18}\orcid{0000-0003-1205-5108}
 Masashi \textsc{Omiya},\altaffilmark{4,5}
 Mayuko \textsc{Mori},\altaffilmark{4,5}\orcid{0000-0003-1368-6593}
 Motohide \textsc{Tamura} \altaffilmark{4,19,20}\orcid{0000-0002-6510-0681}
 Nobuhiko \textsc{Kusakabe},\altaffilmark{21,4,5}\orcid{0000-0001-9194-1268}
 Noriharu \textsc{Watanabe},\altaffilmark{1}\orcid{0000-0002-7522-8195}
 S\'{e}bastien \textsc{Vievard},\altaffilmark{22}\orcid{0000-0003-4018-2569}
 Taiki \textsc{Kagetani},\altaffilmark{1,5}\orcid{0000-0002-5331-6637}
 Takashi \textsc{Kurokawa},\altaffilmark{23}
 Takayuki \textsc{Kotani},\altaffilmark{4,5,6}\orcid{0000-0001-6181-3142}
 Takuma \textsc{Serizawa},\altaffilmark{5,23} \orcid{0009-0009-5823-0793}
 Tomoyuki \textsc{Kudo},\altaffilmark{22}\orcid{0000-0002-9294-1793}
 Vigneshwaran \textsc{Krishnamurthy},\altaffilmark{24}\orcid{0000-0003-2310-9415}
 Yugo \textsc{Kawai},\altaffilmark{1}\orcid{0000-0002-0488-6297}
 and
 Yuya \textsc{Hayashi}\altaffilmark{1}\orcid{0000-0001-8877-0242}
    }

\altaffiltext{1}{\komaba}
\altaffiltext{2}{\komabains}
\altaffiltext{3}{\iac}
\altaffiltext{4}{\abc}
\altaffiltext{5}{\naoj}
\altaffiltext{6}{\sokendai}
\altaffiltext{7}{\ep}
\altaffiltext{8}{Department of Social Data Science, Hitotsubashi University, 2-1 Naka, Kunitachi, Tokyo 186-8601, Japan}
\altaffiltext{9}{Department of Science, National Museum of Nature and Science, 4-1-1 Amakubo, Tsukuba, Ibaraki 305-0005, Japan}
\altaffiltext{10}{Space Data Frontiers Research Center, Fujitsu Research, Fujitsu Limited, 4-1-1 Kamikodanaka, Nakahara, Kawasaki, Kanagawa 211-8588, Japan}
\altaffiltext{11}{Graduate School of Environmental, Life, Natural Science and Technology, Okayama University, 3-1-1 Tsushima-naka, Kita, Okayama 700-8530, Japan}
\altaffiltext{12}{\utops}
\altaffiltext{13}{\gron}
\altaffiltext{14}{\elsi}
\altaffiltext{15}{\isas}
\altaffiltext{16}{\lagna}
\altaffiltext{17}{Okayama Observatory, Kyoto University, 3037-5 Honjo, Kamogata, Asakuchi, Okayama 719-0232, Japan}
\altaffiltext{18}{\ritsumei}
\altaffiltext{19}{\astron}
\altaffiltext{20}{Institute of Laser Engineering, The University of Osaka, 2-6 Yamadaoka, Suita, Osaka 565-0871, Japan}
\altaffiltext{21}{Headquarter for Co-Creation Strategy, National Institutes of Natural Sciences, Tokyo, 105-0001, Japan}
\altaffiltext{22}{\subaru}
\altaffiltext{23}{Institute of Engineering, Tokyo University of Agriculture and Technology, 2-24-26 Nakacho, Koganei, Tokyo, 184-8588, Japan}
\altaffiltext{24}{Trottier Space Institute at McGill, McGill University, 3550 University Street, Montreal, QC H3A 2A7, Canada}

\email{izuru-fukuda@g.ecc.u-tokyo.ac.jp}



\maketitle

\begin{abstract}
Recent large-scale transit surveys conducted by space telescopes such as Kepler and TESS have revealed a vast number of exoplanets, uncovering the diversity of their population. One of the remarkable findings is the presence of a deficiency region in the period–radius distribution of short-period ($<$ 10 days) Neptune-sized planets (4--8\,$R_\oplus$). This region is classified into the Neptune desert ($<$ 3.2 days), the ridge (3.2–5.7 days), and the savanna ($>$ 5.7 days) based on orbital period, each likely reflecting distinct evolutionary pathways.
In this study, we used the InfraRed Doppler (IRD) instrument on the Subaru Telescope to determine the mass of the super-Neptune TOI-1883 b, which resides in the ridge region ($P \sim 4.51~d$) orbiting an M dwarf. 
We measured a planetary mass of $M_p = 13.7^{+6.8}_{-6.5}\,M_\oplus$ and a mean density of $\rho_p = 0.4^{+0.3}_{-0.2}\,\mathrm{g\,cm^{-3}}$, \textcolor{black}{ with $3\sigma$ upper limits of $34.1~M_\oplus$, and $5\sigma$ upper limits of $47.7~M_\oplus$.}
\textcolor{black}{These results suggest that TOI-1883 b is likely a low density super-Neptune.}
We also find that the boundary of the \textit{Neptune desert} defined by planets orbiting FGK-type stars exhibits a similar distribution for planets around M-type stars.
According to the population-based argument of Bourrier et al. (2025), this suggests that TOI-1883 b may have undergone disk-driven migration to reach its current orbit and experienced early atmospheric photoevaporation driven by strong stellar XUV irradiation. 
The derived planetary mass is comparable to or exceeds the conventional critical core mass. We suggest that the high metallicity of the host star ($[\mathrm{Fe/H}] = 0.32 \pm 0.18$) may have suppressed the onset of runaway gas accretion. 
Furthermore, TOI-1883 b has a high Transmission Spectroscopy Metric (TSM > 140), making it an excellent target for future atmospheric characterization via transmission spectroscopy. 
\end{abstract}


\section{Introduction} \label{sec:intro}

Recent transit surveys conducted by space-based missions such as 
\textit{Kepler} \citep{Borucki11} and the Transiting Exoplanet Survey 
Satellite (\textit{TESS}; \cite{Ricker15}) have revealed that a large number of planets exist beyond the Solar System. As the catalog of exoplanets has grown, statistical studies examining where planets reside and what physical properties they exhibit have become increasingly important. One particularly notable feature identified from these surveys is the so-called "\textit{Neptune desert}" \citep{benitez2011mass,szabo2011short,youdin2011exoplanet,beauge2012emerging,lundkvist2016hot, Mazeh16}.
The \textit{Neptune desert} was first identified by \citet{szabo2011short} and was subsequently characterized using \textit{Kepler} data as a statistically significant dearth of Neptune-radius planets $(4\text{--}8\,R_{\oplus})$ on short-period orbits, predominantly around FGK-type stars \citep{Mazeh16}.

More recently, several exceptional planets have been found within this region (e.g., \cite{GJ3470, TOI-674}), leading to a revised interpretation in which the short-period parameter space is subdivided into three regimes based on orbital period: the "\textit{desert}" ($P < 3.2$~days), the ``\textit{ridge}'' (3.2--5.7~days), and the ``\textit{savanna}'' ($P > 5.7$~days) \citep{2024CastroGonzalez}. While the \textit{desert} contains very few planets, the \textit{ridge} shows a localized overdensity, and the \textit{savanna} region transitions to a more diffuse distribution \citep{2024CastroGonzalez,doyle2025exploring}.
The \textit{Neptune desert} is generally interpreted as the outcome of either intense photoevaporation driven by high-energy stellar irradiation \citep{lammer2003atmospheric, vidal2003extended, vidal2004detection,des2007diagram,owen2012planetary,tian2015atmospheric,owen2019atmospheric} or tidal disruption following high-eccentricity migration induced by gravitational perturbations from companion objects \citep{matsakos2016origin,owen2018photoevaporation}, or both.


Planets residing on the \textit{ridge} are thought to have arrived there either through disk-driven migration \citep{goldreich1979towards,lin1986tidal, lin1996orbital, tanaka2002three, baruteau2016formation} or high-eccentricity migration \citep{wu2003planet,ford2008origins,nagasawa2008formation,chatterjee2008dynamical,correia2011tidal,beauge2012multiple}. \citet{bourrier2025atreides} proposed that planetary mean bulk density may serve as a diagnostic to distinguish between these two pathways: planets with densities below $\sim 1\,\mathrm{g\,cm^{-3}}$ are likely shaped by disk-driven migration, whereas those above this threshold may represent planets that survived tidal disruption following high-eccentricity migration {\citep{castro2026neptunian}}. 
Although these formation scenarios have been widely discussed, observational data remain limited, especially for short-period planets around M dwarfs, where the sample size is notably small. Moreover, M dwarfs exhibit stronger and more long-lived XUV emission than solar-type stars \citep{linsky2014radiation, chadney2015xuv, mcdonald2019sub,gaidos2024radius}, potentially shifting the efficiency of atmospheric mass loss and altering the boundaries of the \textit{Neptune desert} \citep{szabo2023sub,magliano2024revisiting}.
Therefore, determining the masses of planets around M dwarfs that lie within the \textit{ridge} region---where the number of confirmed systems remains limited---is crucial for constraining their densities, atmospheric retention, and evolutionary histories. Such measurements provide key insight into the origins of short-period Neptune-sized planets orbiting M dwarfs, and help elucidate the physical processes that shape the observed distribution, including the formation of the \textit{desert} and \textit{ridge} regions.

In this paper, we report on the characterization of a short-period super-Neptune around the mid-M dwarf TOI-1883 (Table \ref{tb:stellar}). The planet, TOI-1883~b, was originally detected as a planetary candidate from the TESS survey and subsequently validated by \citet{pelaez2024validation}.
We characterized the system by follow-up observations with the InfraRed Doppler (IRD) instrument on Subaru Telescope \citep{Tamura12,Kotani18,Kuzuhara18} 
and a series of the Multicolor Simultaneous Camera for studying Atmospheres of Transiting exoplanets (MuSCAT; \cite{muscat}) to determine the stellar and planetary properties, including the planetary mass.
This paper is a part of the intensive programs with the Subaru/IRD, dedicated to follow-up observations of transiting planets around M dwarfs discovered by TESS (e.g., \cite{Hirano20b,Hirano21,Fukui22,Mori22,Kawauchi22,Kagetani23,Hirano23,Barkaoui24,Hori23,ikuta2025mass}).

The rest of this paper is organized as follows. In Section \ref{sec:data}, we describe the data of the photometry for the planetary transits and high resolution spectroscopy for the stellar characterization and radial velocity (RV).
In Section \ref{sec:result}, we derive the stellar and planetary properties both from photometry and spectroscopy. We also analyze potential transit timing variations (TTVs) using multiple transit observations.
In Section \ref{sec:discussion}, we discuss comparisons with other planets in the \textit{Neptune desert} and \textit{ridge}, implications for possible formation pathways, and prospects for future studies.
In Section \ref{sec:conclusion}, we conclude this paper. 

\begin{table}[tb]
\caption{Stellar parameters of TOI-1883}
\begin{center}
\begin{tabular}{lc}
\hline \hline
Parameter & TOI-1883 \\ \hline 
(Literature Values) &  \\
TIC  & 348755728  \\
2MASS ID   & J08562138-1255503  \\
Gaia ID &  5735744144510573696 \\ 
$\alpha$ (J2000)$^{\rm a}$  & 08:56:21.42  \\
$\delta$ (J2000)$^{\rm a}$  &  -12:55:50.32 \\
$\mu_\alpha \cos \delta$ (mas yr$^{-1}$)$^{\rm a}$ &30.780 $\pm$ 0.020\\
$\mu_\delta$ (mas yr$^{-1}$)$^{\rm a}$   & 	2.836 $\pm$ 0.020 \\
parallax (mas)$^{\rm a}$ &   8.5081 $\pm$ 0.0240  \\
Gaia $G$ (mag)$^{\rm a}$  & 14.5019 $\pm$ 0.0003  \\
TESS $T$ (mag)$^{\rm b}$   & 13.3462 $\pm$ 0.0075  \\
$V$ (mag)$^{\rm b}$ & 15.792  $\pm$ 0.069 \\
$J$ (mag)$^{\rm b}$   & 11.871 $\pm$ 0.024 \\
$H$ (mag)$^{\rm b}$   & 11.234 $\pm$ 0.024  \\
$K$ (mag)$^{\rm b}$  & 10.994 $\pm$ 0.023  \\ \hline
(Derived Values) &   \\
$d$ (pc)$^{\rm }$   & $117.538^{+0.335}_{-0.326}$ \\
$T_{\rm eff}$ (K)$^{\rm }$   & $3554^{+26}_{-24}$  \\
$\lbrack{\rm Na/H}\rbrack$ (dex)  & 0.30 $\pm$ 0.16  \\
$\lbrack{\rm Mg/H}\rbrack$ (dex)  & 0.49 $\pm$ 0.26  \\
$\lbrack{\rm Ca/H}\rbrack$ (dex) & 0.39 $\pm$ 0.16   \\
$\lbrack{\rm Ti/H}\rbrack$ (dex) & 0.69 $\pm$ 0.25   \\
$\lbrack{\rm Cr/H}\rbrack$ (dex) & 0.39 $\pm$ 0.14  \\
$\lbrack{\rm Mn/H}\rbrack$ (dex) & 0.55 $\pm$ 0.21  \\
$\lbrack{\rm Fe/H}\rbrack$ (dex) &0.32 $\pm$ 0.18   \\
$\lbrack{\rm Sr/H}\rbrack$ (dex) &  0.47 $\pm$ 0.27  \\
$\lbrack{\rm M/H}\rbrack$ (dex) & 0.45 $\pm$ 0.07  \\
$\log g$ (cgs)$^{\rm }$  & 4.719 $\pm$ 0.018  \\
$M_{\rm s}$ ($M_{\sun}$)   & $0.495 \pm 0.011$  \\ 
$R_{\rm s}$ ($R_{\sun}$)  & $0.506 \pm 0.015$   \\ 
$\rho_{\rm s}$ (g cm$^{-3}$) & $5.41 \pm 0.49$  \\ 
$L_{\rm s}$ ($L_{\sun}$)$^{\rm }$  & $0.0371 \pm 0.0006$  \\  
$v \sin i_{\rm s}$ (km s$^{-1}$)  & < 3  \\ \hline
\end{tabular} \label{tb:stellar}
\end{center}
\begin{tabnote}
\par\noindent
\footnotemark[a] Gaia DR3 \citep{Gaia}
\par\noindent
\footnotemark[b] TESS Input Catalog v8.2 \citep{Stassun19}
\par\noindent
\end{tabnote}
\end{table}

\begin{table}[tb]
\caption{Transit Observations of TOI-1883~b}
\begin{center}
\begin{tabular}{lcc}
\hline \hline
Date & Usage & Filters \\ \hline
(TESS)  & \\
2019/02/02 (Sector 8) & (a) & $T$ \\
2021/02/09 (Sector 35) & (a,b) & $T$ \\
2023/01/18 (Sector 61) & (b) & $T$ \\ \hline
(MuSCAT2) &  \\
2020/11/30 & Clouds & $g, r, i, z_s$ \\
2021/03/14 & (a,b) & $g, r, i, z_s$ \\
2021/03/23 & (a,b) & $g, r, i, z_s$ \\
2022/03/05 & ~~(b)* & $g, r, z_s$ \\
2022/03/23 & Clouds & $g, r, z_s$ \\
\hline
(MuSCAT3) & \\ 
2022/03/09 & ~~(a,b)* & $g, r, i, z_s$ \\ \hline
(Sinistro) & \\
2022/12/10 & (b) & $i_p$ \\ \hline
\end{tabular} \label{tb:transit_obs}

\hfill \parbox{0.4\linewidth}{\footnotesize * Partial transit.}
\end{center}
\begin{tabnote}
\par\noindent
\footnotemark[(a)] \citet{pelaez2024validation}
\par\noindent
\footnotemark[(b)] This study
\end{tabnote}
\end{table}

\section{Observations and Data} \label{sec:data}
After TOI-1883\,b (originally labeled as TOI-1883.01) was validated as a bona-fide planet, we attempted to determine its mass through RV observations using the IRD instrument mounted on the Subaru Telescope. We also conducted a periodogram to check for stellar activity using data from TESS and ASAS-SN (see Appendix \ref{apn:activity_analysis}).
Furthermore, we carried out transit observations with the MuSCAT series and examined potential transit timing variations (TTVs) that could arise from gravitational perturbations induced by an additional outer planet.
Below, we provide details of observations.

\subsection{TESS photometry} \label{sec:tess}
TOI-1883 (TIC 348755728) was observed by TESS in Sectors 8, 35, and 61. A transit signal with a period of $\sim4.506$ days was first identified in the Sector 8 data through the analysis of full-frame images (FFIs) using the MIT Quick-Look Pipeline (QLP;\citet{huang2020photometry}; \citet{kunimoto2021quick},\citet{kunimoto2022quick}) and the faint target transit search. 
Furthermore, \citet{pelaez2024validation} validated the planetary nature of TOI-1883.01 from the first two sectors (8, 35) with the Gaia DR3 catalog \citep{Gaia} and the TOI catalog \citep{Guerrero21} by incorporating follow-up observations with ground-based reconnaissance photometry through the TESS Follow-up Observing Program (TFOP; \cite{Collins19}).

\begin{figure}[tb]
\begin{center}
\includegraphics[width=8cm]{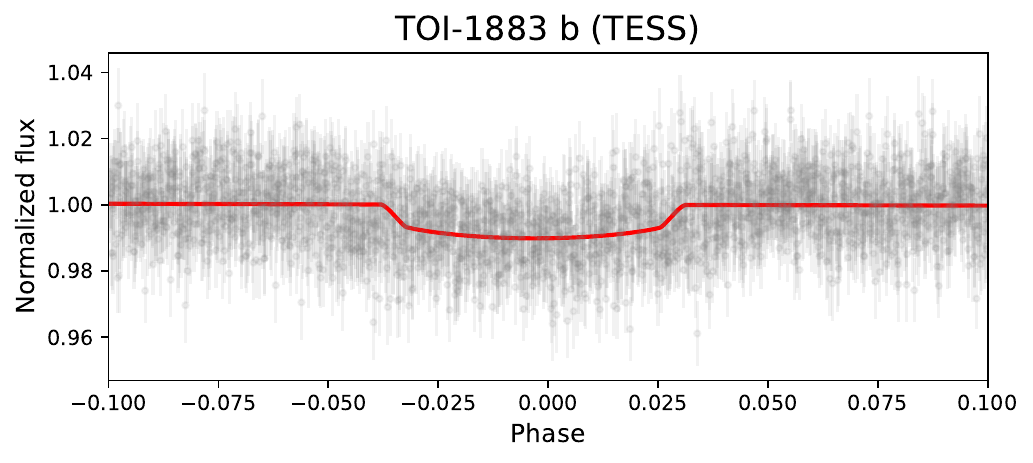}
\end{center}
\caption{Phase-folded TESS light curve (gray). The derived orbital period (= 4.506 days) and the optimum transit model (red) are shown within transit windows spanning three times the transit duration near the transit center. {Alt text: Time-series flux for the transit of the TESS data and the optimum model.} 
} \label{fig:tess}
\end{figure}

We extracted the Presearch Data Conditioning Simple Aperture Photometry (PDC-SAP) for two sectors (35 and 61) from the Mikulski Archive for Space Telescopes (MAST). Since no TESS data processed by the Science Processing Operations Center (SPOC) pipeline \citep{Jenkins16} were available for Sector 8, it was excluded from the present analysis.
We remove data with bad-quality flags from each sector and use the data within transit windows with a width of three times the predicted transit duration ($\sim$ 2.04 hours) (Figure \ref{fig:tess}). The fluxes are normalized by the median value of the data out of the transit window for each sector, where there is no significant flux variability attributed to stellar activity or systematics (see Section \ref{sec:activity}).

\subsection{Ground-based photometry -- TCS 1.52m / MuSCAT2} \label{sec:m2}
\begin{figure*}[tb]
\begin{center}
\includegraphics[width=8cm]{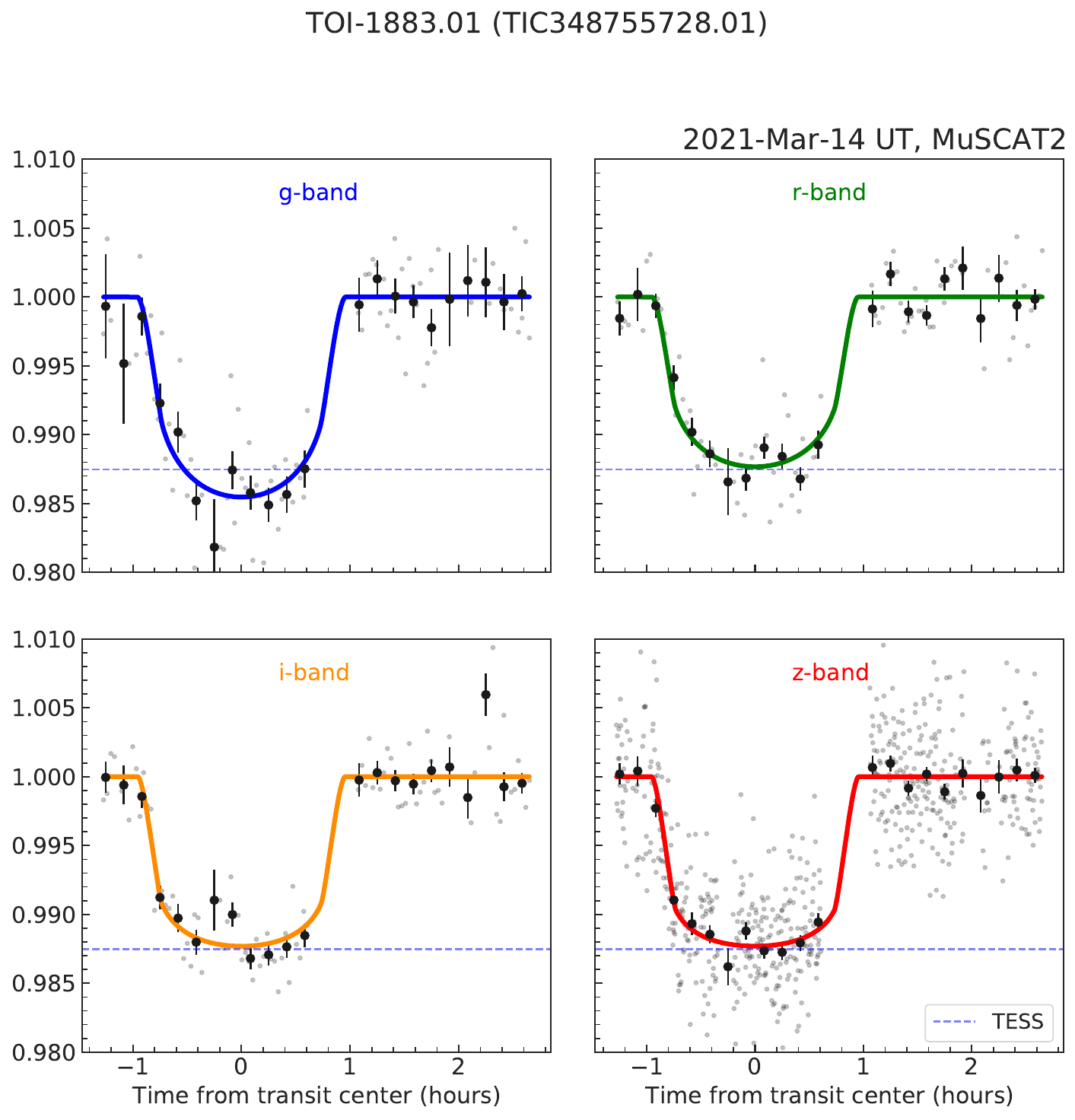}
\includegraphics[width=8cm]{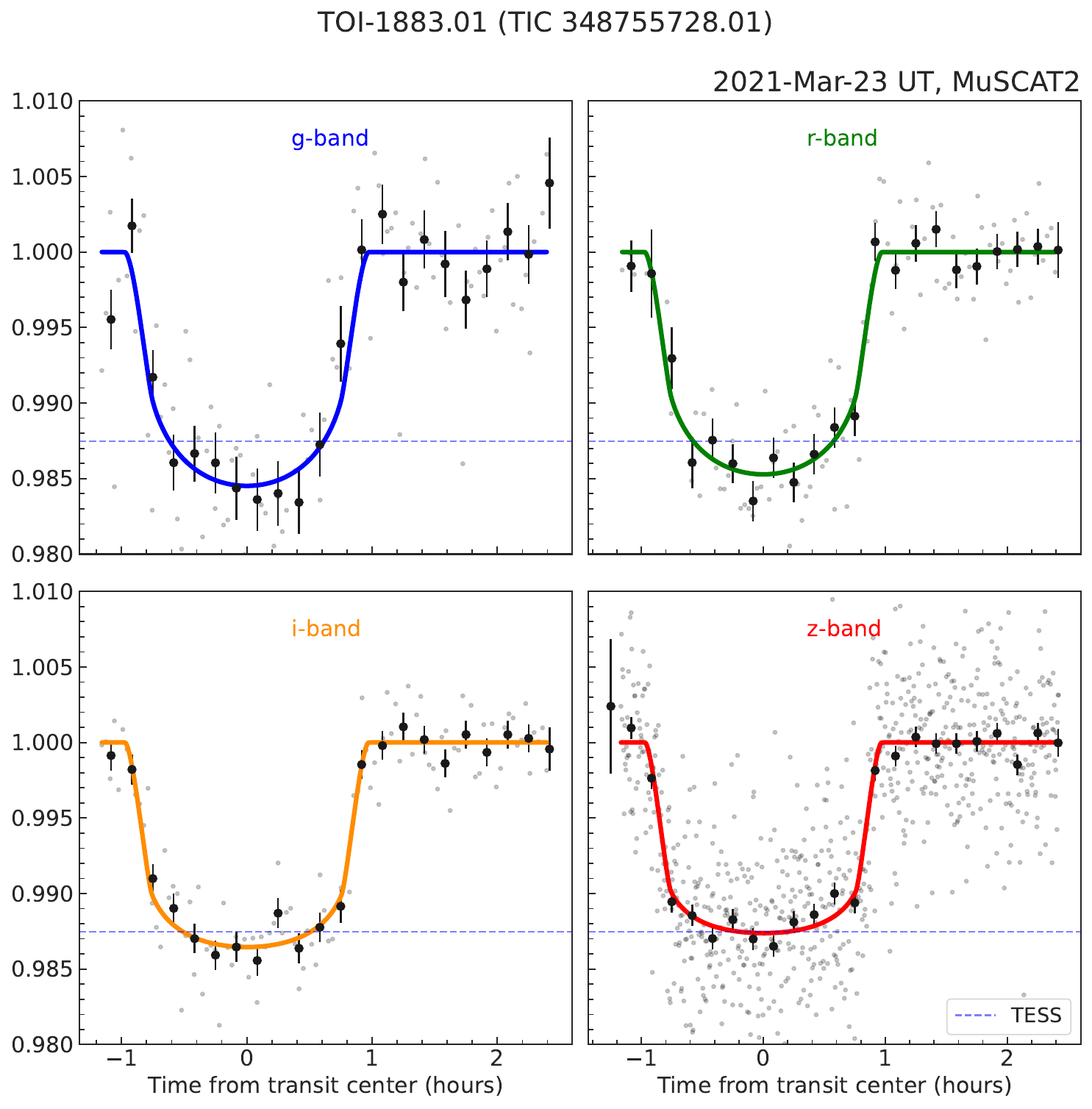}
\includegraphics[width=8cm]{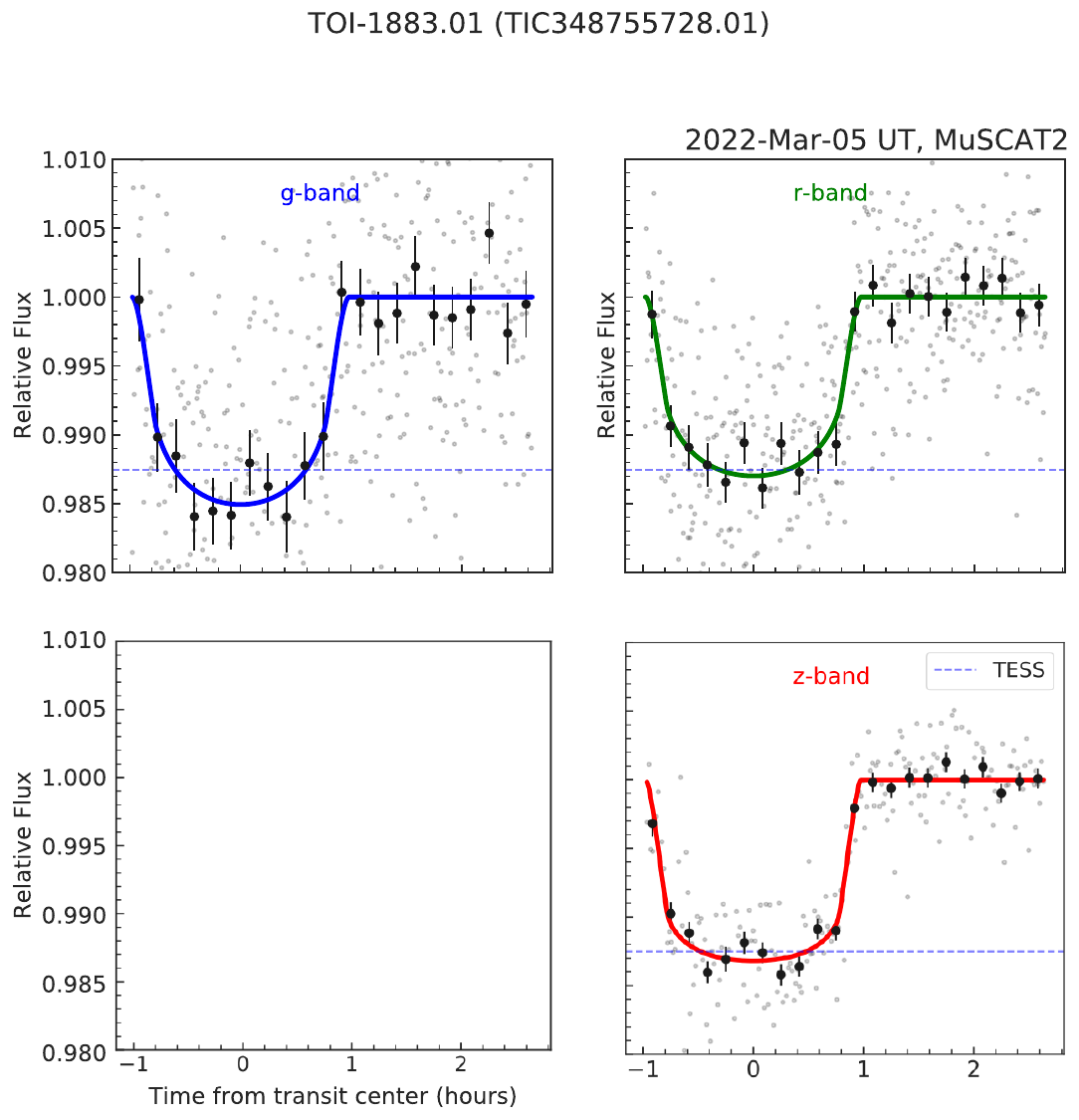}
\includegraphics[width=8cm]{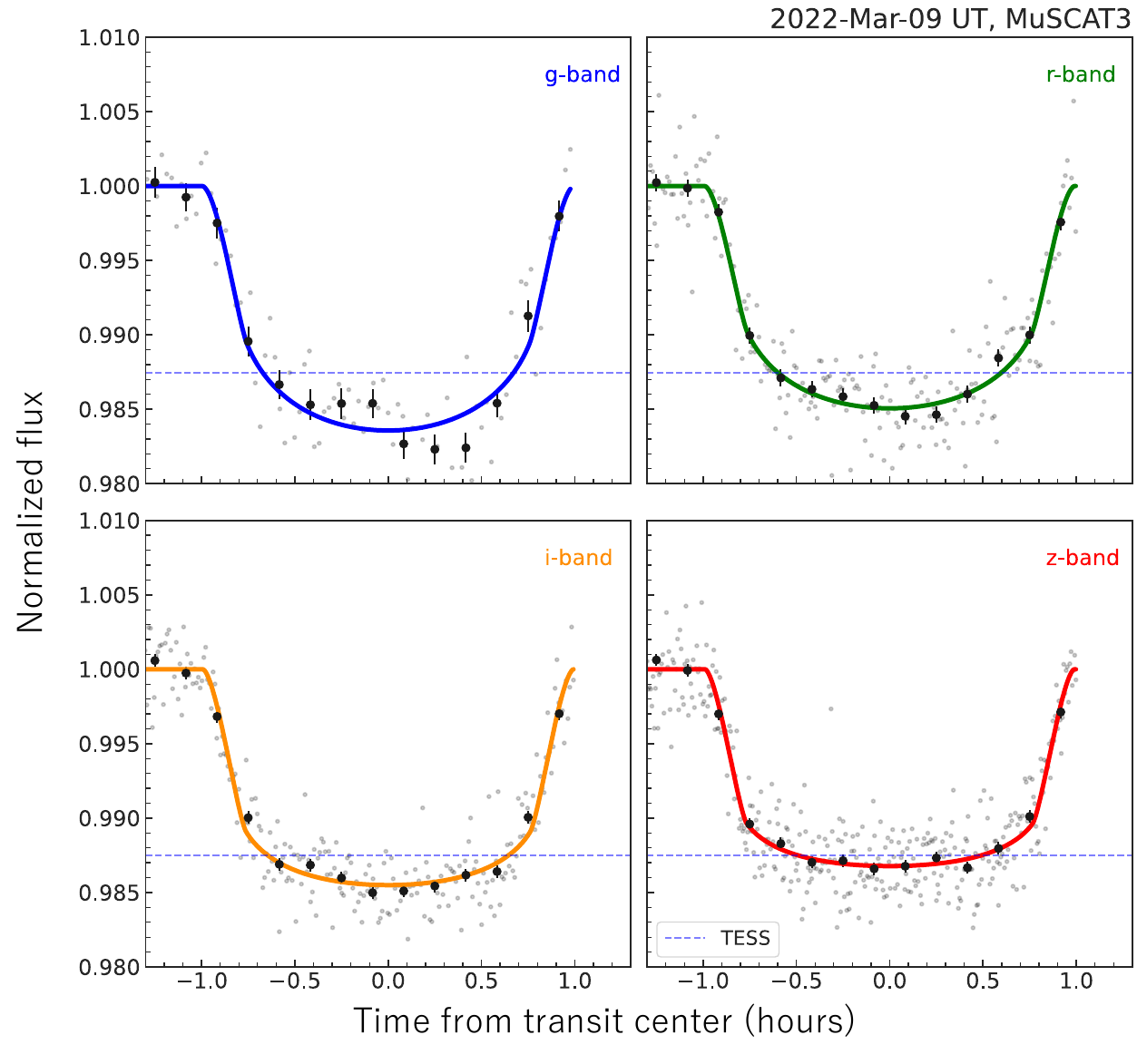}
\end{center}
\caption{Multicolor simultaneous light curves in $g$-, $r$-, $i$-, and $z_s$-bands, obtained by MuSCAT2 on 2021 March 14, 2021 March 23, 2022 March 5, and MuSCAT3 on 2022 March 9 (Section \ref{sec:m2} and \ref{sec:m3}). For MuSCAT3, only a partial transit was observed. The data are jointly fitted with the transit models (blue, green, orange, and red) and the baseline model, and the observed fluxes (gray) are binned into points shown in black. However, on 2022 March 5, the $i$-band was not operational, and only three bands were available.
{Alt text: Time-series flux for the transits of the MuSCAT2 and MuSCAT3 data in $g$-, $r$-, $i$-, and $z_s$-bands, and the optimum models for each.} 
}
\label{fig:muscat}
\end{figure*}

We conducted follow-up transit observations of TOI-1883~b five times from 2020 November 30 to 2022 March 23 with the multi-band simultaneous camera MuSCAT2 \citep{muscat2}, mounted on the 1.52m Carlos S\'{a}nchez telescope (TCS) at the Teide Observatory in the Canary Islands, Spain (Table \ref{tb:transit_obs}). MuSCAT2 has four optical channels, each of which is equipped with a 1k$\times$1k CCD camera with a pixel scale of 0".44 pixel$^{-1}$, and is capable of simultaneous imaging in $g$-, $r$-, $i$-, and $z_s$-bands with a field of view of 7'4$\times$7'4. 
The transit light curves observed with MuSCAT2 are presented in Figure \ref{fig:muscat}, except for the observations on 2020 November 30 UTC and 2022 March 23, which were severely affected by clouds and therefore excluded. Additionally, on March 5, 2022 UTC, due to a malfunction of the $i$-band camera, the observations are available in only the remaining three bands. The nominal exposure times are $120$, $120$, $120$, and $15s$ in the $g$-, $r$-, $i$-, and $z_\mathrm{s}$- bands, respectively. Due to weather conditions, on 2022 March 5, we used $30\,\mathrm{s}$ exposures in the $g$-, $r$-, and $z$- bands.
We calibrate the images and perform aperture photometry to extract relative photometry by following the procedure described in \citet{Fukui16} and application to MuSCAT2 data (e.g., \cite{Fukui21,Hayashi24}) using a dedicated pipeline developed by \citet{Fukui11}. 

\begin{figure}[tb]
\begin{center}
\includegraphics[width=8cm]{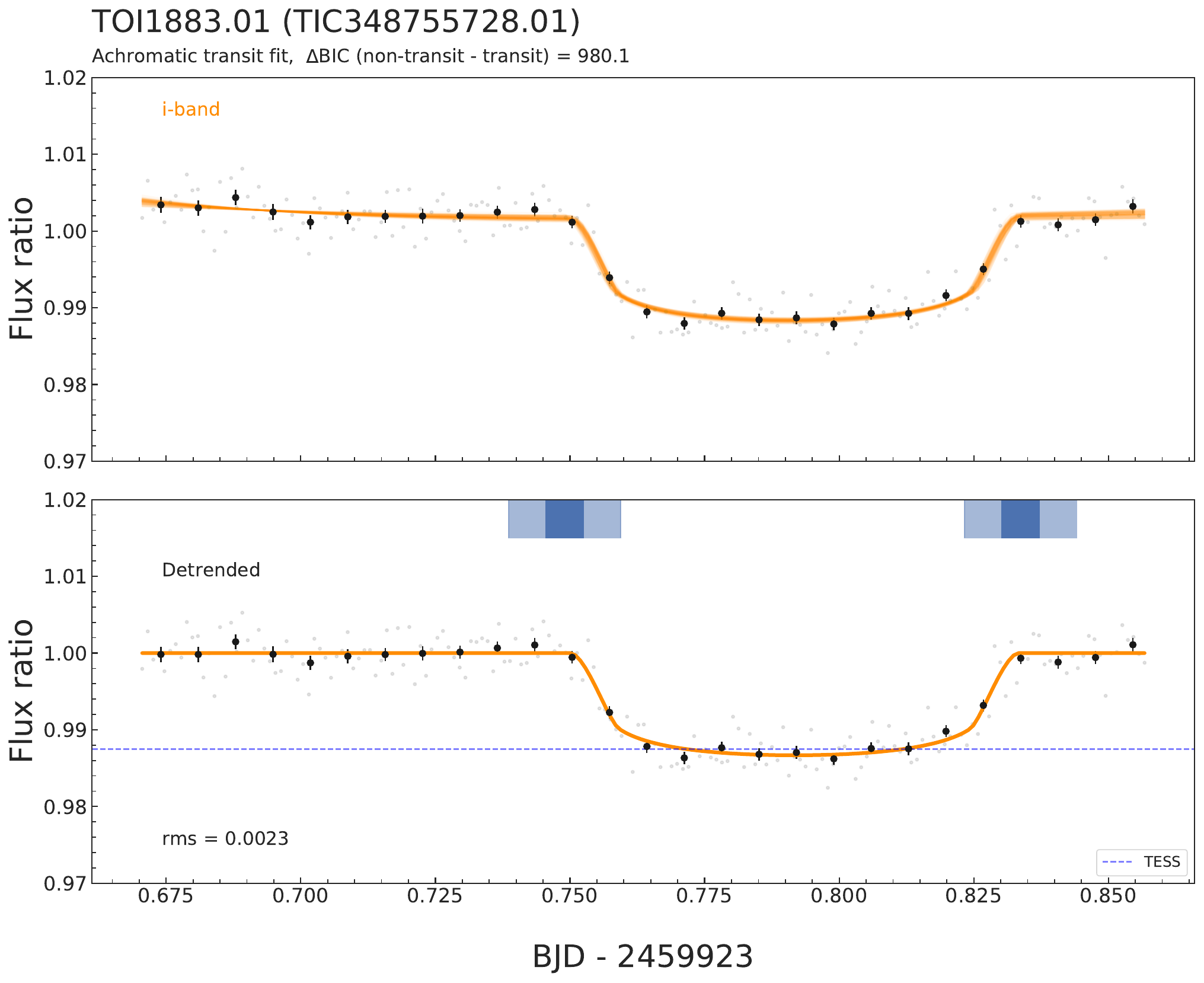}
\end{center}
\caption{This figure shows the transit light curve of TOI-1883.01 observed with the $i$-band of LCO/Sinistro. The upper and lower panels present the data before and after the detrending, respectively. 
The orange line represents the model light curve, the gray points indicate the observed flux, and the black dots show the binned data. The predicted ingress and egress times from the TESS ephemeris are marked in the lower panel by the blue and light-blue vertical lines, respectively.
{Alt text: Time-series flux for the transits of the Sinistro data in $i$-band, and the optimum model.} 
}
\label{fig:sinistro}
\end{figure}

\subsection{Ground-based photometry -- FTN 2m / MuSCAT3} \label{sec:m3}
We conducted a follow-up transit observation of TOI-1883~b once with the multi-band simultaneous camera MuSCAT3 \citep{muscat3}, mounted on the 2m Faulkes Telescope North (FTN) at Haleakala Observatory in Hawaii, the United States (Table \ref{tb:transit_obs}). MuSCAT3 has four optical channels, each of which is equipped with a 2k$\times$2k CCD camera with a pixel scale of 0".266 pixel$^{-1}$, and is capable of simultaneous imaging in $g$-, $r$-, $i$-, and $z_s$-bands with a field of view of 9'1 $\times$ 9'1. The exposure times were set to be 90, 30, 35, and 20 s in $g$-, $r$-, $i$-, and $z_s$-bands, respectively.
The raw images are processed with \texttt{BANZAI} pipeline \citep{banzai}, and the aperture photometry is conducted in the same way for MuSCAT2 (Section \ref{sec:m2}). The light curve observed with MuSCAT3 is shown in Figure \ref{fig:muscat}.

\subsection{Ground-based photometry -- LCO 1m / Sinistro} \label{sec:sinistro}
We conducted a follow-up transit observation of TOI-1883~b on 2022 March 9 with Sinistro (Table \ref{tb:transit_obs}), an optical camera mounted on one of the 1m telescopes located at McDonald Observatory in Texas, the United States, operated by Las Cumbres Observatory \citep{brown2013cumbres}. Sinistro has a 26'.5 $\times$ 26'.5 field of view with a pixel scale of 0.389''. 
We observed the target for a total of 400 minutes in the $i_p$-band, with an exposure time of 60 min. The data were reduced by the standard LCOGT \texttt{BANZAI} pipeline \citep{banzai}, and photometry was performed using \texttt{AstroImageJ} software \citep{collins2017astroimagej}. The light curve observed with Sinistro is shown in Figure \ref{fig:sinistro}.

\begin{figure}[tb]
\begin{center}
\includegraphics[width=8cm]{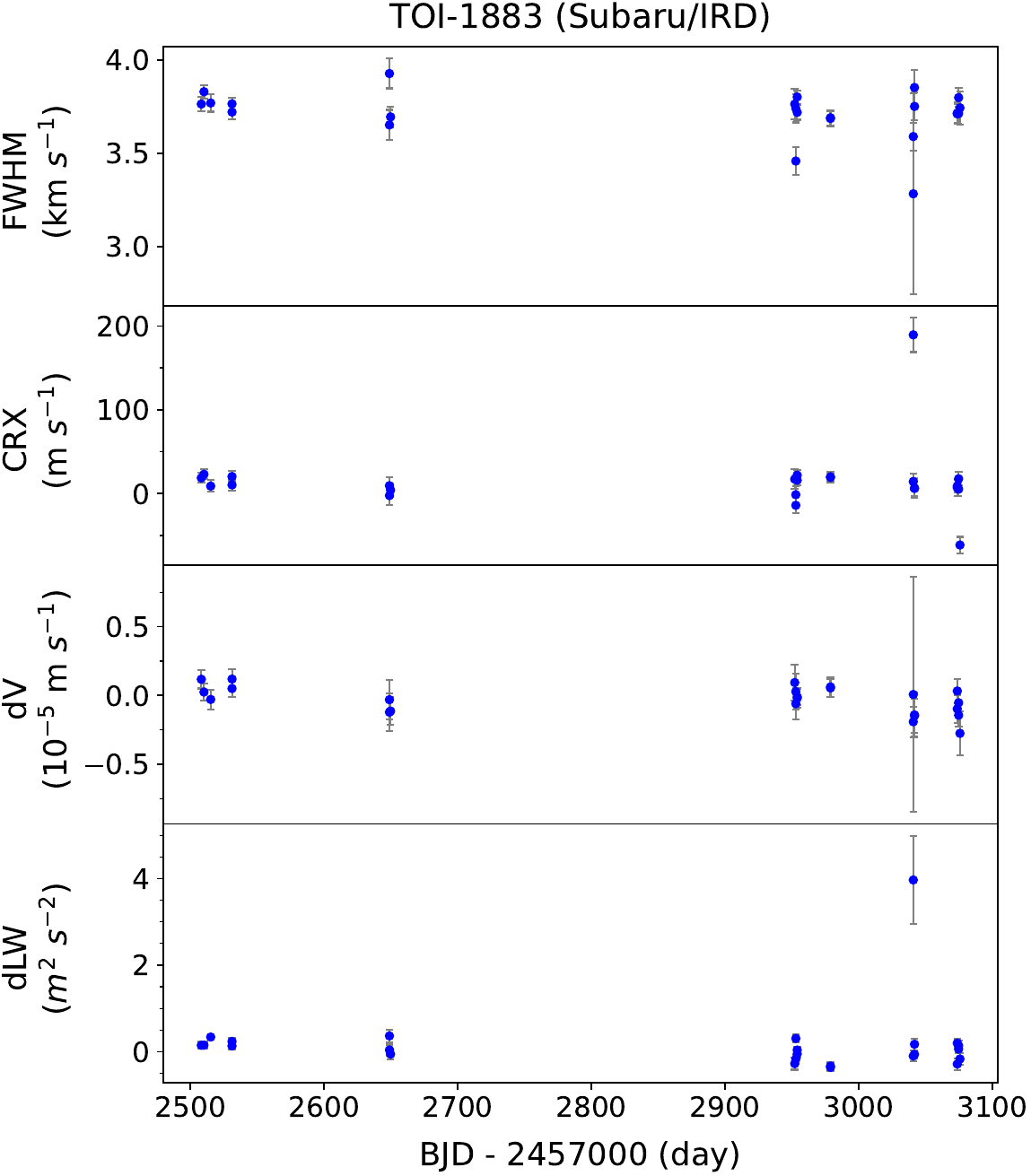}
\end{center}
\caption{FWHM, CRX, dV, and dLW (blue) for TOI-1883 from the IRD spectra (Section \ref{sec:ird}). 
{Alt text: FWHM, CRX, dV, and dLW, from the IRD spectra for each panel.
} 
} \label{fig:rv}
\end{figure}

\subsection{High resolution spectroscopy -- Subaru 8.2m / IRD} \label{sec:ird}
We obtained time-series of high-resolution spectra of TOI-1883 with the IRD instrument \citep{Tamura12, Kotani18, Kuzuhara18} between UT 2021 October 19 and 2023 May 10, under the Subaru-IRD TESS intensive follow-up programs (ID: S21B-118I, S23A-067I; PI: Norio Narita). 
The IRD is a high-precision, high-dispersion (R $\approx$ 70,000) NIR spectrograph mounted on the Subaru 8.2 m telescope. The wavelength range is 930 - 1740 nm, which covers the $Y$-, $J$-, and $H$- bands, with two detectors covering the $YJ$- and $H$-bands, respectively. The exposure time for all data was set at 1800 s to get a high signal-to-noise ratio (S/N). 1-D extracted spectra have S/N ratios of 14.9 - 27.6 per pixel at 1000 nm and 14.6 - 52.6 per pixel at 1600 nm. We obtained 37 spectra with an S/N from 17 to 90 per pixel, and the raw data are reduced by the procedure in \citet{Kuzuhara18}, \citet{Hirano20}, and \citet{Kuzuhara24}.
Briefly, 1-D spectra are extracted from the observed raw data and fitted with the stellar template spectrum processed from the individual spectra. The spectra are divided into segments, and the RV is calculated from each segment for each exposure. Also, telluric lines are removed from the stellar spectrum. For more information on data reduction, see \citet{Hirano20} and \citet{Kuzuhara24}.

From January 7th to 23rd, 2022, the detector associated with the $YJ$-bands experienced technical issues, resulting in the inability to conduct observations in this wavelength range. Consequently, there are 13 more data points in the $H$-band only. 

Although we initially obtained 37 IRD measurements, some data were affected by poor weather conditions (e.g., clouds passing that degraded the adaptive optics performance) or by accidentally targeting the companion star located $\sim$15 arcsec away. After excluding these measurements, 29 measurements are left, which we use for the subsequent analysis. Considering the overall data quality, we adopted the integrated spectra over the wavelength range corresponding to the $H$- band as the primary dataset to ensure the robustness and reliability of our analysis (Table \ref{tb:rv_data}).



The activity indicators, the full width at half maximum (FWHM; the line width), BiGauss (dV; the line asymmetry) by fitting Gaussian functions \citep{Santerne15}, chromatic index (CRX; the wavelength dependence of RV) and the differential line width (dLW; the line width) \citep{Zechmeister18}, are also computed to investigate the stellar activity as in \citet{Harakawa22} (Figure \ref{fig:rv}).

\begin{table}[tb]
\caption{Relative RVs of TOI-1883 for $H$- band}
\begin{center}
\begin{tabular}{c c c c}
\hline \hline
BJD & RV ($\text{m/s}$) & $\sigma$ ($\text{m/s}$) & S/N \\ \hline
    2459508.13195 & 18.58 & 5.50 & 25  \\
    2459510.12698 & -2.72 & 5.83 & 45  \\
    2459515.13797 & 0.29 & 6.63 & 38  \\
    2459531.09264 & 14.83 & 5.80 & 43  \\
    2459531.11390 & 8.25 & 6.28 & 42 \\
    2459588.06860 & -0.01 & 7.79 & $78^{*}$ \\ 
    2459588.08999 & -4.83 & 8.76 & $75^{*}$ \\
    2459589.00806 & 11.6 & 7.29 & $89^{*} $ \\
    2459589.02937 & 27.84 & 7.02 & $89^{*} $ \\
    2459589.05070 & 13.21 & 7.80 & $90^{*} $ \\
    2459589.07205 & 4.41 & 8.07 & $90^{*} $ \\
    2459598.05062 & 20.15 & 6.80 & $83^{*}$ \\
    2459598.07200 & 8.63 & 8.18 & $83^{*} $ \\
    2459598.09341 & 0.00 & 8.37 & $82^{*} $ \\
    2459602.05766 & -1.86 & 13.95 & $36^{*} $ \\
    2459602.07908 & 36.88 & 11.44 & $38^{*}$ \\
    2459604.02311 & -20.10 & 8.75 & $71^{*} $ \\
    2459604.04446 & -16.02 & 9.32 & $69^{*} $ \\
    2459952.11571 & -16.90 & 8.11 & 23 \\
    2459952.93369 & -19.99 & 7.21 & 27  \\
    2459952.95497 & -16.13 & 6.94 & 28  \\
    2459953.95728 & -26.52 & 5.89 & 42  \\
    2459953.97856 & -15.36 & 6.03 & 43  \\
    2460041.73693 & 12.93 & 8.70 & 24 \\
    2460041.77999 & 7.87 & 7.41 & 30  \\
    2460073.74035 & -9.05 & 6.88 & 27  \\
    2460073.76568 & -22.60 & 7.21 & 34  \\
    2460074.75466 & 1.26 & 7.40 & 33  \\
    2460074.77594 & -8.33 & 6.80 & 33  \\
    \hline
\end{tabular} \label{tb:rv_data}
\end{center}
\begin{tabnote}
\par\noindent
\footnotemark[*] Due to technical issues with the cameras related to the $Y$- and $J$-bands, observations are limited to the $H$-band only, resulting in S/N recorded at 1600 nm. All other S/N values are recorded at 1000 nm.
\end{tabnote}
\end{table}

\section{Analyses and Results} \label{sec:result}

\subsection{Stellar properties}
\subsubsection{Spectroscopic properties} \label{sec:metal}
We estimated the metallicity ([Fe/H]) and effective temperature ($T_{\mathrm{eff}}$) of the host star using the IRD spectra. 
In the derivation, we compared the equivalent widths (EWs) of atomic and molecular absorption lines between the observed and theoretical spectra.

To determine $T_{\mathrm{eff}}$, we used 47 FeH molecular lines located in the wavelength range of 990-1012~nm (see \citet{Ishikawa22} for details).  
We also measured the elemental abundances using a total of 25 atomic lines corresponding to Na~\textsc{i}, Mg~\textsc{i}, Ca~\textsc{i}, Ti~\textsc{i}, Cr~\textsc{i}, Mn~\textsc{i}, Fe~\textsc{i}, and Sr~\textsc{ii} \citep{Ishikawa20}.  
The surface gravity (log~$g$) and microturbulent velocity were fixed at 5.0 and 0.50~km~s$^{-1}$, respectively.  
We confirmed that the effects of fixing these parameters on the derived values of $T_{\mathrm{eff}}$ and [Fe/H] are negligibly small.
First, we derived $T_{\mathrm{eff}}$ from the equivalent widths of the FeH lines under the assumption of solar composition.  
Next, using the derived $T_{\mathrm{eff}}$, we determined the abundances [M/H] for the eight elements mentioned above.  
From this analysis, we obtained $T_{\mathrm{eff}} = 3554^{+26}_{-24}\,\mathrm{K}$ and $\mathrm{[Fe/H]} = 0.32 \pm 0.18$ dex.
The final results obtained are summarized in Table \ref{tb:stellar}.

\begin{figure}[tb]
\begin{center}
\includegraphics[width=8cm]{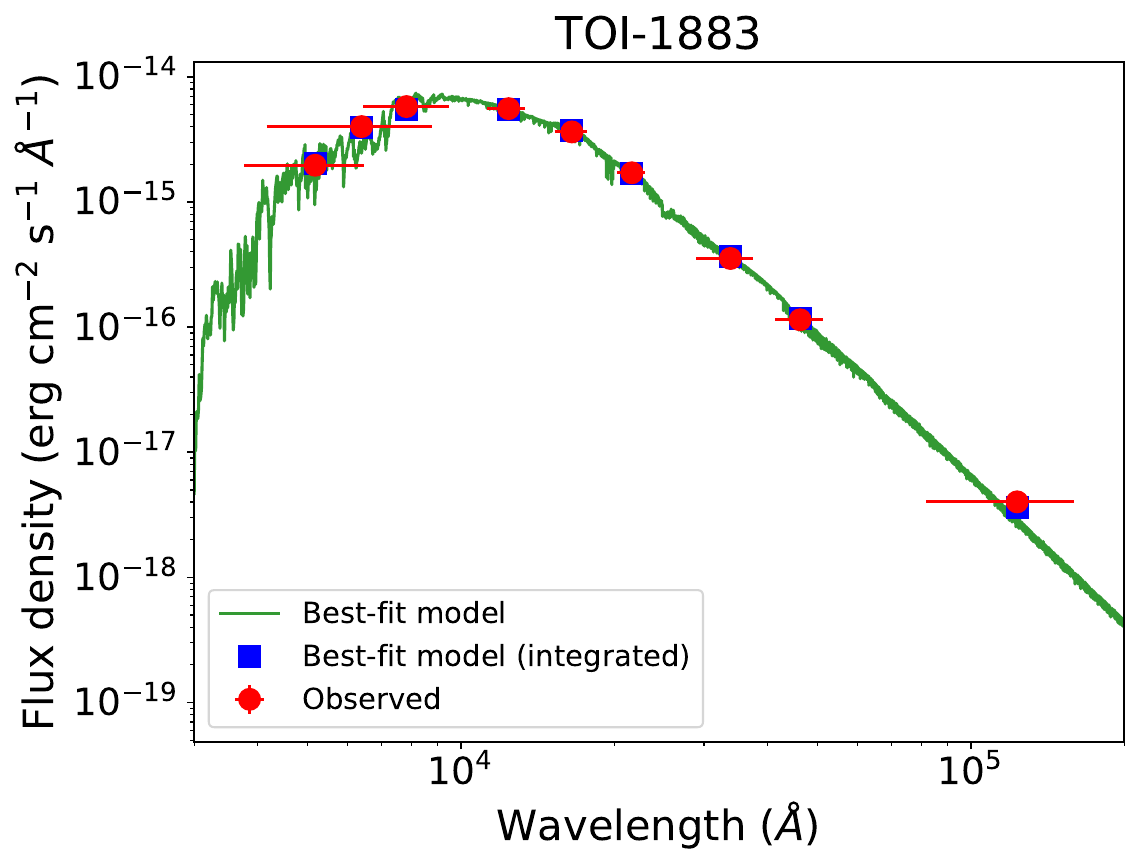}
\end{center}
\caption{Spectrum energy distributions (SED) of TOI-1883. Green curve shows the best-fit SED model. Red diamonds and blue squares are the data and integrated best-fit model for the photometric bands, respectively.
{Alt text: Wavelength (\AA) versus flux density (erg cm$^{-2}$ s$^{-1}$ {\rm \AA}$^{-1})$} as the SED. Diamond and square show the observed data and optimum point. 
Solid line shows the best-fit SED model.
} \label{fig:sed}
\end{figure}

\begin{figure}[tb]
\begin{center}
\includegraphics[width=8cm]{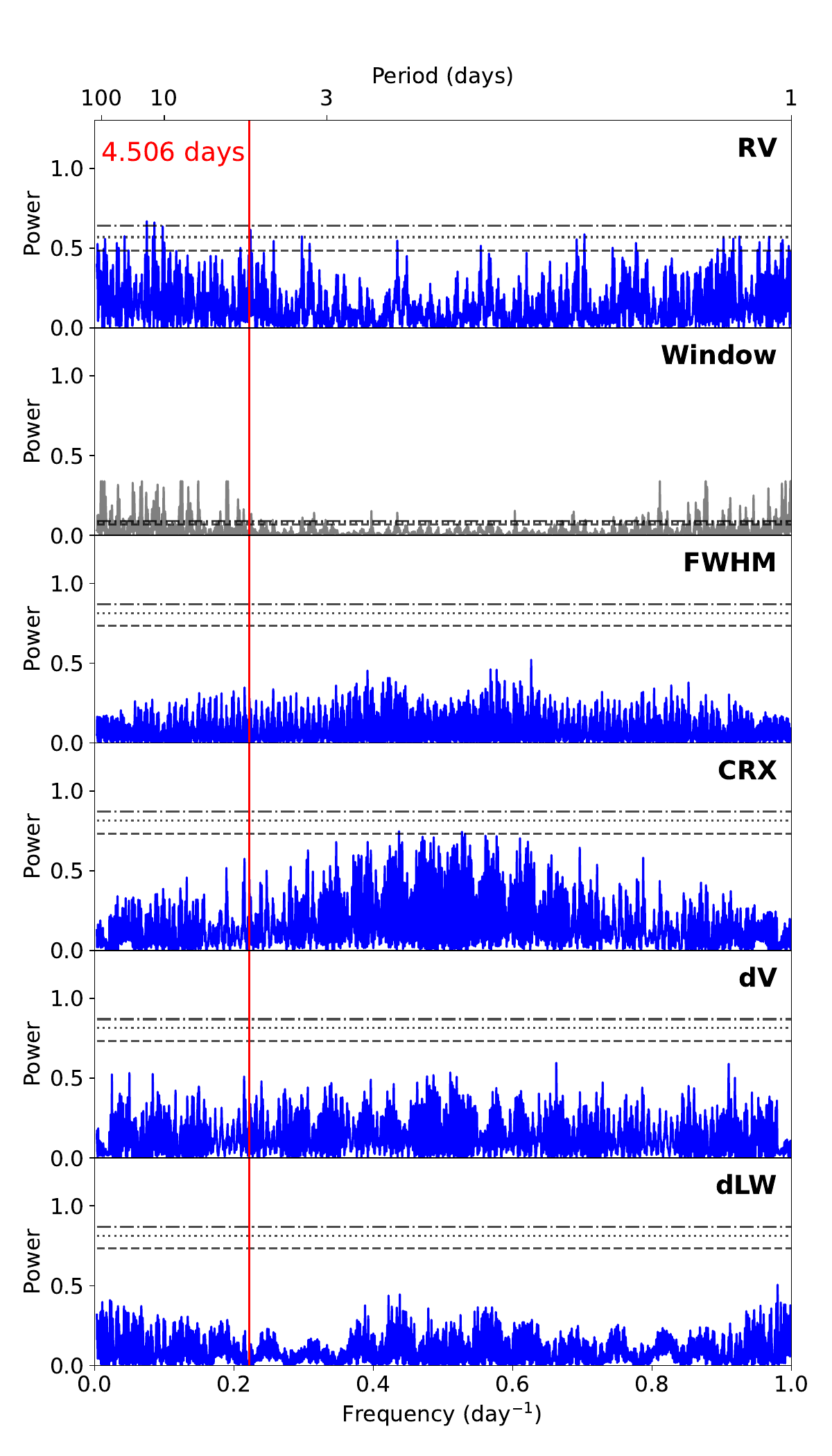}
\end{center}
\caption{Periodograms of the RV, Window, FWHM, CRX, dV, and dLW, obtained with the GLS from the IRD spectra for TOI-1883 (Section \ref{sec:activity}).
The horizontal lines represent the FAP of 0.10, 1.0, and 10.0 \%, respectively (black) for each of the panels.
{Alt text: Period (day) versus the power of the GLS periodogram for the RV, Window, O-C, FWHM, dV, CRX, and dLW, from the IRD spectra. In each panel, the vertical line shows the orbital period of the transiting planet, and the horizontal lines show the thresholds of the FAP of 0.10, 1.0, and 10.0 \%.} 
} \label{fig:periodogram}
\end{figure}

\subsubsection{Photometric properties} \label{sec:sed}
Apart from the metallicity [Fe/H], the other stellar parameters were estimated from the photometric properties.  
Using the parallax from \textit{Gaia}~DR3 \citep{gaia2016, gaia2023} (assuming the distance $d$ as the inverse of the parallax), the $K_s$-band magnitude from the Two Micron All Sky Survey (2MASS; \citep{skrutskie2006}), and the [Fe/H] value derived in the previous section, we estimated the stellar radius ($R_{\mathrm{s}}$) and mass ($M_{\mathrm{s}}$) based on the empirical luminosity–metallicity–radius and luminosity–metallicity–mass relations presented by \citet{Man15, Man19}.  
As a result, we obtained $M_{\mathrm{s}} = 0.495 \pm 0.011$ $M_{\sun}$ and $R_{\mathrm{s}} = 0.506 \pm 0.015$ $R_{\sun}$.

Subsequently, we updated $T_{\mathrm{eff}}$, $R_{\mathrm{s}}$, surface gravity (log~$g$), and luminosity ($L_s$) by fitting the stellar atmosphere models to the observed spectral energy distribution (SED) (Figure \ref{fig:sed}).  
For more details, see \citet{ikuta2025mass}.  
In this analysis, the IRD-derived $T_{\mathrm{eff}}$ and [M/H] were adopted as priors for the posterior probability estimation of the parameters.  
As a result, we obtained $T_{\mathrm{eff}} = 3554^{+26}_{-24}\,\mathrm{K}$, 
$R_{\mathrm{s}} = 0.509^{+0.009}_{-0.008}\,R_{\odot}$, 
$\log g = 4.719 \pm 0.018$, 
and $L = 0.0371 \pm 0.0006\,L_{\odot}$.
However, because the stellar radius ($R_{\mathrm{s}}$) derived from the synthetic atmosphere fitting may be affected by model-dependent uncertainties, we adopted the empirically determined values of $R_{\mathrm{s}}$ and $M_{\mathrm{s}}$ derived in the preceding paragraph as the final stellar radius and mass ($M_{\rm s}$ $=0.495 \pm 0.011$ $M_{\sun}$, $R_{\rm s}$ $=0.506 \pm 0.015$ $R_{\sun}$). 
The surface gravity (log~$g$) was then recalculated based on these adopted values.
In addition, we independently drew random values of the stellar mass $M_{\mathrm{s}}$ and radius $R_{\mathrm{s}}$ from their respective normal distributions and calculated the stellar density, obtaining $\rho_{\mathrm{s}} = 5.41 \pm 0.49$ g cm$^{-3}$ (Table \ref{tb:stellar}).

\subsubsection{Period analysis and stellar activity} \label{sec:activity}
We perform period analysis of the RV, Window, activity indicators, FWHM, CRX, dV, and dLW (Section \ref{sec:ird}) with the Generalized Lomb-Scargle periodogram (GLS; \cite{GLS}). 
\textcolor{black}{The False-Alarm Probability (FAP) levels were calculated using the analytic prescription of GLS, as implemented in the PyAstronomy GLS routine. We used the `powerLevel` method to obtain the power thresholds corresponding to FAPs of 10\%, 1\%, and 0.1\% over the searched frequency range.}
As shown in Figure~\ref{fig:periodogram}, no periodic signal corresponding to the planet (P = 4.506~days) was detected in the RV.
Three signals are found at periods between 10--30 days that marginally exceed the FAP of 0.1\%. 
However, all of them show strong overlap with the window function, indicating that these peaks are likely influenced by the observational sampling. Therefore, we do not regard these periodogram peaks alone as independent evidence for a planetary signal.
In addition, no significant periodic signals were detected in any of the stellar activity indicators.
In the case of M-type stars, the contribution of macroturbulent velocity and other broadening components to the rotation kernel becomes significant. Still, their exact values are not well constrained, making accurate modeling difficult. Nevertheless, there is no indication of rapid stellar rotation, and the projected rotational velocity is estimated to be $ v\sin i < 3~\mathrm{km~s^{-1}}$ from the IRD spectra. 

\begin{figure}[tb]
\begin{center}
\includegraphics[width=8cm]{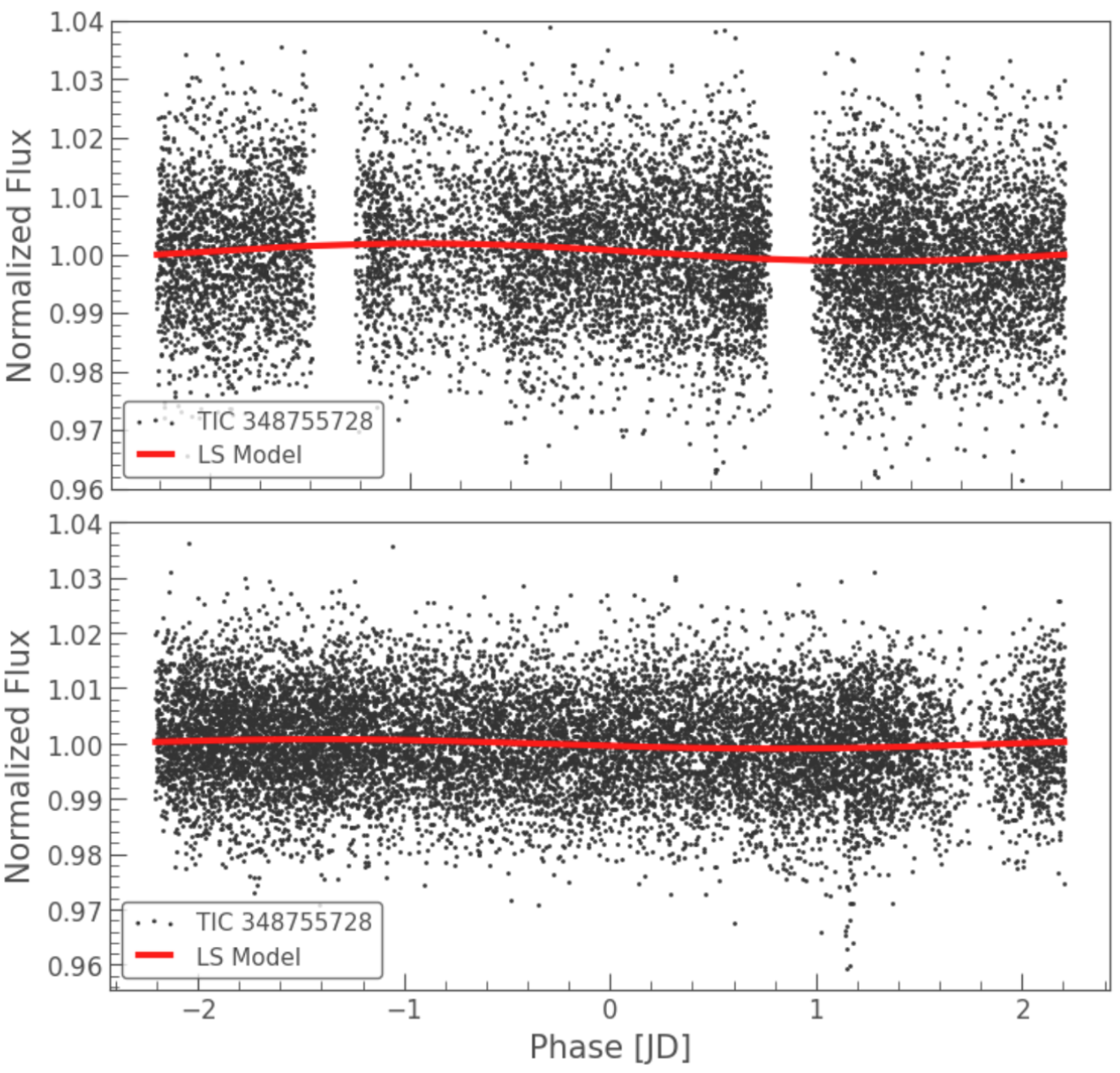}
\end{center}
\caption{
Phase-folded at a period of 4.508~days, transit-masked, and normalized TESS flux (black) together with the best-fit light-curve model (red). The panels from top to bottom show the data from Sectors~35, 61, respectively (only the observations with an exposure time of 120~s are used). TIC~348755728 corresponds to TOI-1883. “LS Model” denotes the Lomb–Scargle light curve model.
{Alt text: The TESS flux is shown after masking the transit signals and phase-folding the light curve.}
} \label{fig:stellar activity}
\end{figure}

To investigate the stellar activity of the host star, we search for photometric variability in the TESS light curve.
We performed a GLS periodogram analysis of the TESS light curve after masking the planetary transits and found no significant periodic signals, with the variability amplitude being as small as $\sim$ 0.10~\% (Figure \ref{fig:stellar activity}).
Although an excess of power is observed at long periods, the corresponding period lies well beyond the observational time baseline and is therefore poorly constrained.
Such low-frequency signals are more likely attributable to sampling-related systematics rather than to a planetary companion.
Accordingly, we do not consider this feature further in this study.
In addition, we search for periodic variability from the light curves of the All-Sky Automated Survey for Supernovae (ASAS-SN; \cite{asas}).
We retrieve the ASAS-SN $V$-filter light curves from the ASAS-SN portal \citep{Hart23}, spanning from approximately October 2013 to November 2018. We calculate a periodogram from the light curves after removing outliers. As a result, we also find no significant periodicities of photometric variability due to stellar activity. The ASAS-SN light curve and periodogram are shown in Appendix \ref{apn:activity_analysis} (Figure \ref{fig:asas-sn_lc} and Figure \ref{fig:asas-sn_period}).

\subsection{Planetary properties} \label{sec:planet}

\subsubsection{Transit model for photometry} \label{sec:transit}
We modeled the phase-folded light curve TESS, MuSCAT2, MuSCAT3, and Sinistro of TOI-1883 \,b, and inferred the posterior distributions of the transit and baseline parameters using an MCMC sampler with \texttt{emcee} \citep{emcee}. 
The light curve was first folded on the known orbital period $P$ at epoch $T_{0,\mathrm{BJD}}$ \citep{2024PelaezTorres}, and we restricted the analysis to the phase range $-0.10 \le \phi \le 0.10$ to focus on the transit. Unless otherwise noted, we used the unbinned data
(we tested a weak binning option with 200--500 bins only for performance checks).
If per-point flux uncertainties were unavailable or non-finite, we robustly
estimated a common photometric scatter from the out-of-transit (OOT) region
($|\phi|>0.060$) via the median absolute deviation:
$\sigma_{\mathrm{OOT}} \sim 1.5\,\mathrm{median}\!\left(\,|f-\mathrm{median}(f)|\,\right)$,
and adopted it as a constant error for those points.

We employed the quadratic limb-darkening law parameterized by the
\citet{kipping13} re-parameterization $(q_1,q_2)\in[0,1]^2$, which maps to
the physical coefficients $(u_1,u_2)$ via
$u_1 = 2\sqrt{q_1}\,q_2$ and $u_2 = \sqrt{q_1}(1-2q_2)$.
The transit light-curve model was computed using the \texttt{batman} package \citep{kreidberg2015batman}, which predicts the flux $F_{\mathrm{tr}}(\phi \,|\, r_p, a/R_\star, i, u_1,u_2)$, where $r_p\!=\!R_p/R_\star$, $a/R_\star$ is the scaled semi-major axis, and $i$ is
the orbital inclination. 
For the transit-shape analysis, the forward model included a small global phase shift $\Delta\phi$, equivalent to a refinement of the reference epoch $T_0$, together with a linear baseline and an additive white-noise jitter. For the TTV analysis, no global phase shift was applied; instead, the mid-transit times of individual transits were fitted independently.



We adopted uniform or weakly informative priors consistent with physically
allowed ranges used in the code (Table \ref{tb:transit_prior}). The likelihood assumes independent Gaussian errors:
\begin{equation}
\ln \mathcal{L} \;=\; -\frac{1}{2}\sum_k \left[
\frac{\bigl(f_k - y(\phi_k)\bigr)^2}{\sigma_k^2} + \ln\!\bigl(2\pi \sigma_k^2\bigr)
\right].
\end{equation}
where the index $k$ runs over all observational data points.

We initialized the sampler around a plausible starting point
$(r_p, a/R_\star, i, q_1, q_2, \log\sigma_{\mathrm{jit}}, \Delta\phi)$ using small
Gaussian perturbations. We ran \texttt{emcee} with 32 walkers for 5{,}000 burn-in
steps, and then drew 20{,}000 production steps, monitoring
convergence through visual inspection of chains and the stability of summary statistics. A representative maximum-posterior (or median-parameter) model is shown against the phase-folded data in Figure~\ref{fig:tess}, where we plot the best-fit transit model (red) and the TESS data (gray) with the same phase window described above. The phase-folded light curves and the best-fit transit models for the MuSCAT2, MuSCAT3, and SINISTRO observations are presented in the Figure \ref{fig:muscat} and \ref{fig:sinistro}. Furthermore, all of the transit results were utilized to investigate the presence or absence of transit timing variations (TTVs) (Section \ref{sec:ttv}).

\subsubsection{Orbital model for radial velocity} \label{sec:rv}
To constrain the planetary mass, we carried out an RV fit with a circular, 1-planet model that includes five parameters: the RV semi-amplitude $K$, the orbital period $P$, transit central time $T_{\rm c}$, the RV offset $V_{\rm 0}$ and the systematics with the jitter term $\sigma_{\rm jit}$.
The orbital period $P$ and the transit central time $T_{\mathrm{c}}$ were constrained from the transit observations (see Section \ref{sec:transit}), which we used as normal priors for these parameters. For the other parameters, we assigned uniform prior distributions. The parameter boundaries of the prior distributions are shown in Table \ref{tb:prior_1883}.
In the analysis, we first performed a maximum-likelihood estimation of the 
model parameters using \texttt{scipy.optimize} to obtain the parameter values 
that maximize the likelihood. Subsequently, we carried out an MCMC analysis with \texttt{emcee} \citep{emcee}. We employed 32 walkers and performed 50,000 steps, discarding the initial 10,000 steps as burn-in and excluding them from the posterior distributions.

As a result, we obtained a semi-amplitude of 
$K =8.50 ^{+4.22}_{-3.99}\,\mathrm{m\,s^{-1}}$, which corresponds to a planetary mass of $13.7^{+6.8}_{-6.5}\,M_{\oplus}$. The planetary mass was derived using the stellar mass estimated from empirical relations presented by \citet{Man15, Man19}.
This indicates that the significance level of the detection is approximately at the $2\sigma$ level. 
We also derive the 3$\sigma$ and 5$\sigma$ upper limits as $ 34.1\,M_{\oplus}$ and $ 47.7\,M_{\oplus}$, respectively.
The corner plot of the 1-planet RV model is shown in Figure \ref{fig:corner} (Appendix \ref{apn:1-p RV corner}).
We also tested a Keplerian model incorporating the orbital eccentricity $e$ and the argument of periastron $\omega$. The comparison favored the circular model over the eccentric model ($\Delta \mathrm{BIC} = 3$), and we therefore adopt the circular model as our baseline model in this study.


\begin{table}[tb]
\caption{Prior of the parameters of the transit model}
\begin{center}
\begin{tabular}{lc}
\hline \hline
Parameter & Prior \\
\hline
$R_p/R_\star$ & $\mathcal{U}(0.005,\, 0.3)$ \\
$a/R_\star$ & $\mathcal{U}(5,\, 50)$ \\
$i$ (deg) & $\mathcal{U}(80,\, 90)$ \\
$q_1$ & $\mathcal{U}(0,\, 1)$ \\
$q_2$ & $\mathcal{U}(0,\, 1)$ \\
$c$ (baseline offset) & $\mathcal{U}(0.5,\, 1.5)$ \\
$\log\sigma_{\mathrm{jit}}$ & $\mathcal{U}(-20,\, 0)$ \\
$\Delta\phi$ & $\mathcal{U}(-0.01,\, 0.01)$ \\
$m$ (baseline slope) & $\mathcal{U}(-0.2,\, 0.2)$ \\
\hline
\end{tabular} \label{tb:transit_prior}
\end{center}
\end{table}


\begin{table}[tb]
\caption{Priors of the parameters of the 1-planet RV model}
\begin{center}
\begin{tabular}{lc}
\hline \hline
Parameter & Circular \\ \hline
$K$ ($m/s$) & $\mathcal{U}(0, \infty)$ \\
$P$ ($days$) & $\mathcal{N}(4.5063, 0.0010)$ \\
$tc$ ($BJD$) & $\mathcal{N}(2459256.8461, 0.0028)$ \\
$V_0$ ($m/s$) & $\mathcal{U}(-50, 500)$ \\
$\sigma_{jitter}$ ($m/s$) & $\mathcal{U}(0, 500)$ \\
$\sqrt{e} \sin \omega$ & 0 (fixed) \\
$\sqrt{e} \cos \omega$ & 0 (fixed) \\
$e$ & 0 (fixed) \\
\hline
\end{tabular} \label{tb:prior_1883}
\end{center}
\end{table}

\begin{figure}[tb]
\begin{center}
\includegraphics[width=8cm]{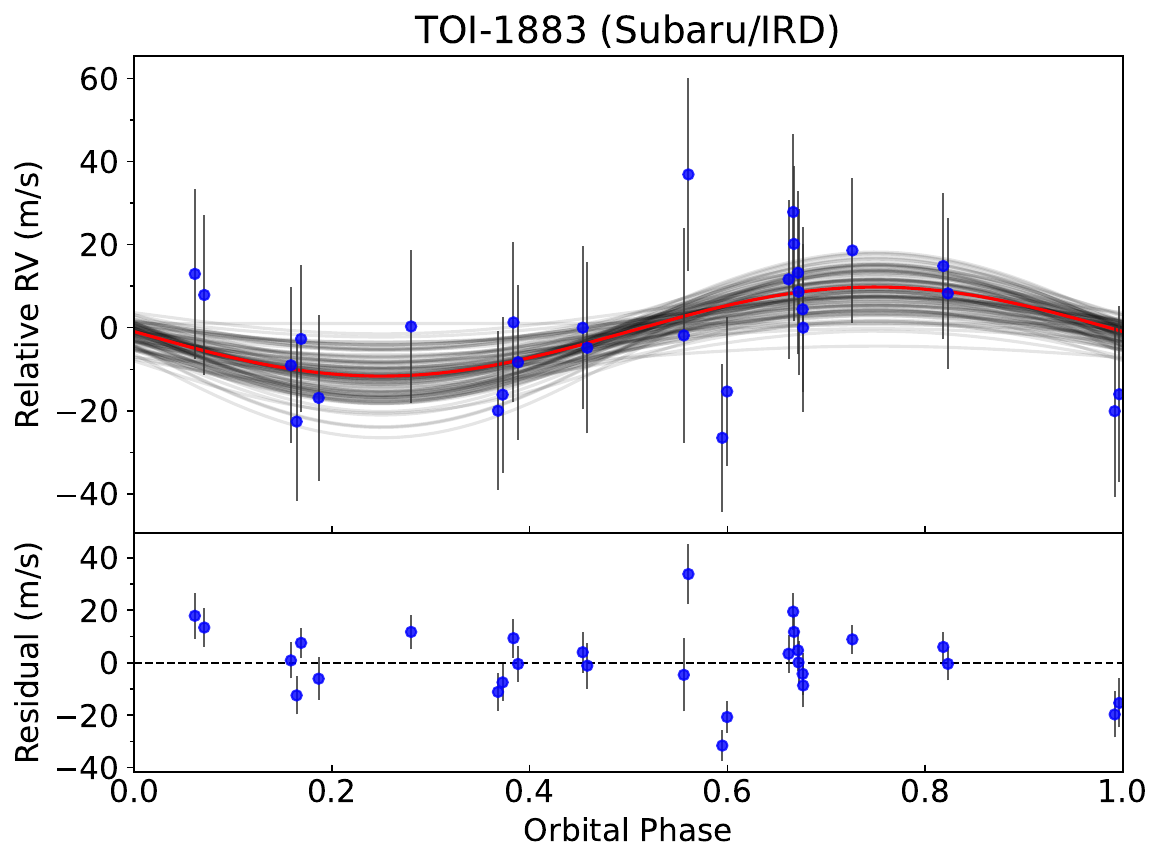}
\end{center}
\caption{Phase-folded radial velocity curve of TOI-1883 observed with Subaru/IRD. The red line shows the best-fit Keplerian model, and the blue points represent the measured RVs with 1$\sigma$ error bars. Thin gray curves indicate 100 model realizations randomly drawn from the posterior samples, illustrating the uncertainty in the fitted model. The bottom panel shows the residuals after subtracting the best-fit model, with an rms scatter of $15.6\,\mathrm{m\,s^{-1}}$.
{Alt text: The orbital phase versus the relative RV (m s$^{-1}$) and residual (m s$^{-1}$). The relative RV is the RV minus offset, and the residual is the relative RV minus the optimum circular model.  The lines show the circular model and its uncertainty.}
} 
\label{fig:1-p RV 1883}
\end{figure}

Figure~\ref{fig:1-p RV 1883} shows the phase-folded RV curve for the 1-planet, circular orbit model, together with the residuals between the observations and the best-fit model.

\begin{table}[tb]
\caption{Parameters of TOI-1883~b}
\begin{center}
\begin{tabular}{lcc}
\hline \hline
Parameter & Value & Reference \\ \hline
(Orbital parameters) & & \\
$P$ (day) &$ 4.508^{+ 0.002 }_{- 0.003 }$ & (1) \\
$T_{\rm c}$ (BJD-2457000) & $ 2256.846 \pm 0.008 $  & (1) \\
$\sqrt{e} \cos \omega$  &0 ({\rm fixed}) & (1) \\
$\sqrt{e} \sin \omega$  & 0 ({\rm fixed}) & (1)  \\
$K$ (m s$^{-1}$) &$ 8.50 ^{+ 4.22 }_{- 3.99 }$ & (1)\\ 
$v_0$ (m s$^{-1}$) &$ -2.65 \pm 3.14$ & (1) \\ 
$\sigma_{\rm jit}$ (m s$^{-1}$)  & $ 13.60 ^{+ 3.00 }_{- 2.50 }$  & (1) \\ \hline
$\log {\cal Z}$ & $-111.718$ & (1) \\ \hline 
(Derived parameters) &   &   \\
$R_{\rm p}/R_{\rm s}$  &$ 0.1076 ^{+ 0.0042 }_{- 0.0016 }$ & (2) \\
$a/R_{\rm s}$  &$ 19.11 ^{+ 0.39 }_{- 0.87 }$ & (2) \\
$R_{\rm p}$ ($R_\earth$)&$  5.65\pm0.24$ & (1) \\
$M_{\rm p}$ ($M_\earth$)  &$ 13.7^{+6.8}_{-6.5}$ & (1) \\
\textcolor{black}{$3\sigma$ upper limit on $M_{\rm p}$ ($M_\oplus$)} & \textcolor{black}{$34.1$} & (1) \\
\textcolor{black}{$5\sigma$ upper limit on $M_{\rm p}$ ($M_\oplus$)} & \textcolor{black}{$47.7$} & (1) \\
$a$ (au) &$0.0423^{+ 0.0010 }_{- 0.0021 }$ & (1) \\
$e$ &$0$ ({\rm fixed})   & (1)  \\
$\omega$ (deg) &$0$ ({\rm fixed}) & (1)  \\
$i_{\rm p}$ (deg) &$89.37^{+ 0.42 }_{- 0.51}$ & (2)     \\
$b$  &$ 0.21 ^{+ 0.15}_{- 0.14}$ & (2)  \\
$T_{14}$ (hour) & $1.951 ^{+ 0.057 }_{- 0.084 }$ & (1) \\
$\rho_{\rm p}$ (g cm$^{-3}$)  &$0.4 ^{+ 0.3 }_{- 0.2}$ & (1) \\ 
$\rho_{\rm p}$ ($\rho_{\earth}$)  &$0.076 ^{+ 0.053 }_{- 0.040}$ & (1) \\ 
$T_{\rm eq}$ (K) ($A_B=0.0$)  &$544 \pm 28 $ & (1) \\
$T_{\rm eq}$ (K) ($A_B=0.3$)  &$ 498 \pm 26$ & (1) \\
$S$ ($S_{\earth}$) &$ 14.7\pm 3.1$ & (1) \\ 
TSM &$143\pm74$ & (1) \\ 
ESM & $10\pm1$ & (1) \\ \hline
\end{tabular}  \label{tb:para1883}
\end{center}
\begin{tabnote}
\par\noindent (1) This work
\par\noindent (2) \citet{2024PelaezTorres}
\end{tabnote}
\end{table}

\subsubsection{Verification of Transit Timing Variations (TTVs)} \label{sec:ttv}
\begin{figure}[tb]
\begin{center}
\includegraphics[width=8cm]{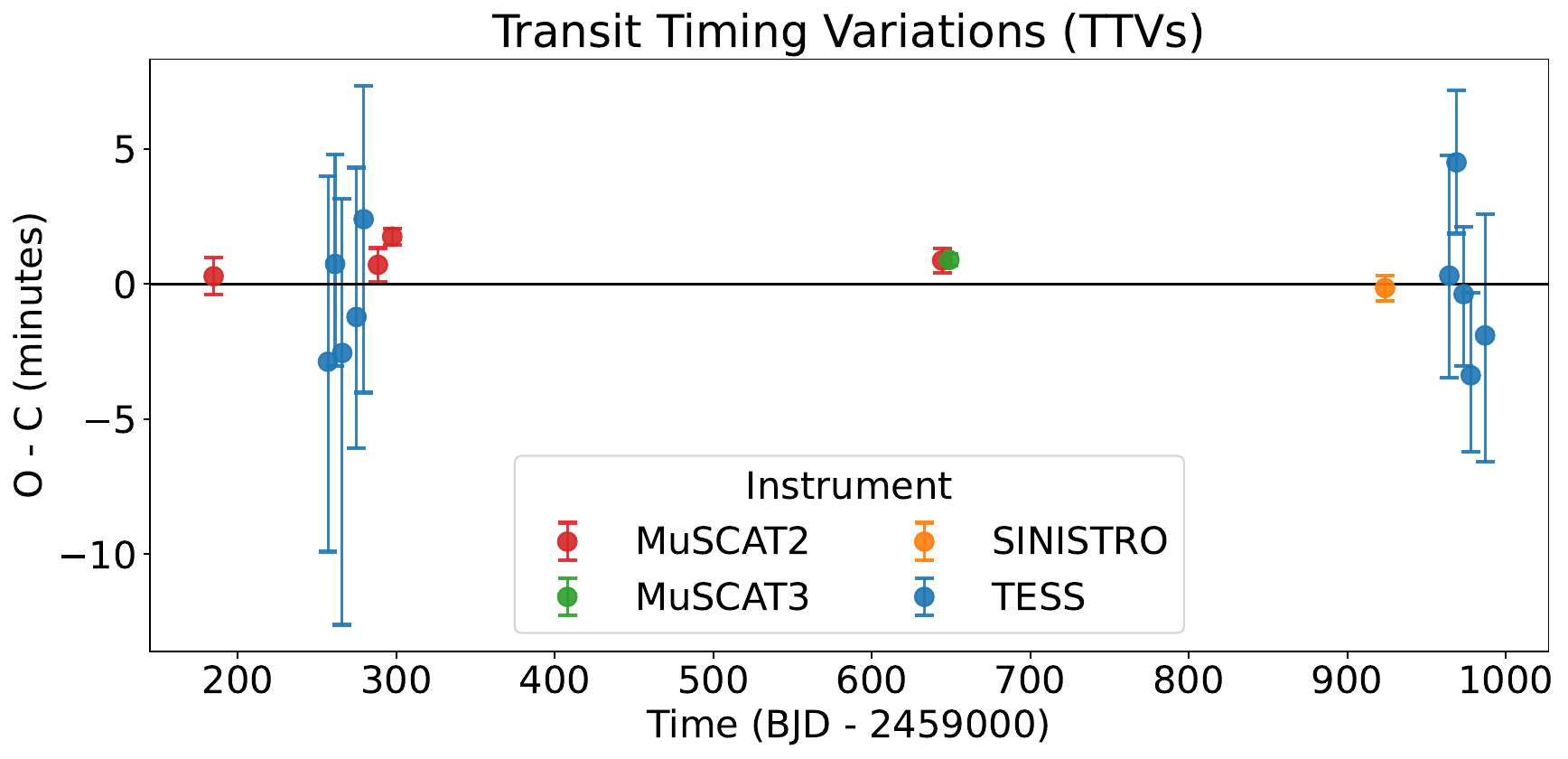}
\end{center}
\caption{
We plot the time series of the residuals between the observed transit central times and the best-fit linear ephemeris. The black line represents the residuals from the linear ephemeris. The colors correspond to the observing instruments.
{Alt text: We plot the time series of the residuals between the observed
transit central times and the best-fit linear ephemeris.}
} 
\label{fig:ttv}
\end{figure}

Given the low precision in the planetary mass determination from the IRD data, the results of the stellar GLS periodogram, and the possibility that the 1-planet model solution may be biased in the presence of an additional planet, we examined the possibility of transit timing variations (TTVs) as a precaution.
We used the transit central time observed with TESS, MuSCAT2, MuSCAT3, and SINISTRO.
Additionally, we utilized the \texttt{allesfitter} package \citep{Gunter2021} to conduct the TTV analysis of the TESS data.
From Figure \ref{fig:ttv}, we find no evidence for the presence of an additional planet. Although the TESS data have relatively large uncertainties, the ground-based observations obtained with MuSCAT2, MuSCAT3, and SINISTRO show much smaller errors, and their residuals are smaller than $\pm ~2$ minutes. Therefore, in the following analysis, we assume that there are no additional planets in this system that would significantly affect the ephemeris residuals of TOI-1883 b.

\begin{figure}[tb]
\begin{center}
\includegraphics[width=9cm]{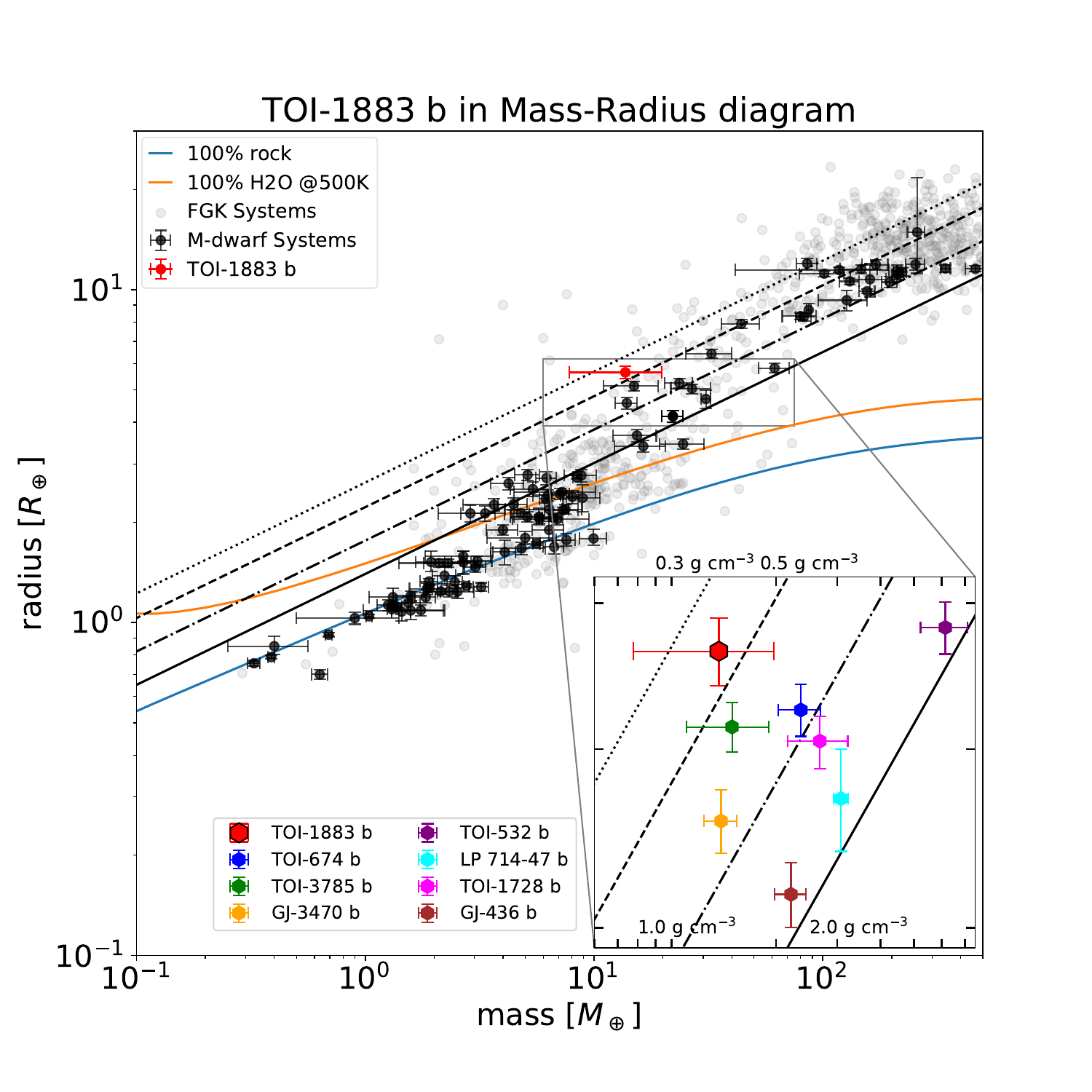}
\end{center}
\caption{
The mass--radius diagram of TOI-1883\,b and other known exoplanets. 
The red point represents TOI-1883\,b, black points show planets orbiting M dwarfs ($T_\mathrm{eff} < 4000$\,$\mathrm{K}$), and gray points show those around FGK-type stars ($4000$\,$\mathrm{K}$ $<$ $T_\mathrm{eff}$ $<$ $6500$\,$\mathrm{K}$). Planets with mean densities comparable to that of TOI-1883 b and radii in the range of 4–6\,$R_{\oplus}$ are color-coded and plotted.
Only the data with upper and lower uncertainties both less than 33~\% in mass and radius are shown. The dotted, dashed, and dash-dotted lines represent mean densities of 0.3, 0.5, and 1\,$\mathrm{g\,cm^{-3}}$ from top to bottom, and the solid curves show theoretical models from \citet{Zeng19}.
{Alt text: The diagram of the mass ($M_\earth$) versus radius ($R_\earth$) for small planets around M dwarfs. The lines show various composition models.}
} \label{fig:MR_1883}
\end{figure}

\begin{figure}[tb]
\begin{center}
\includegraphics[width=8cm]{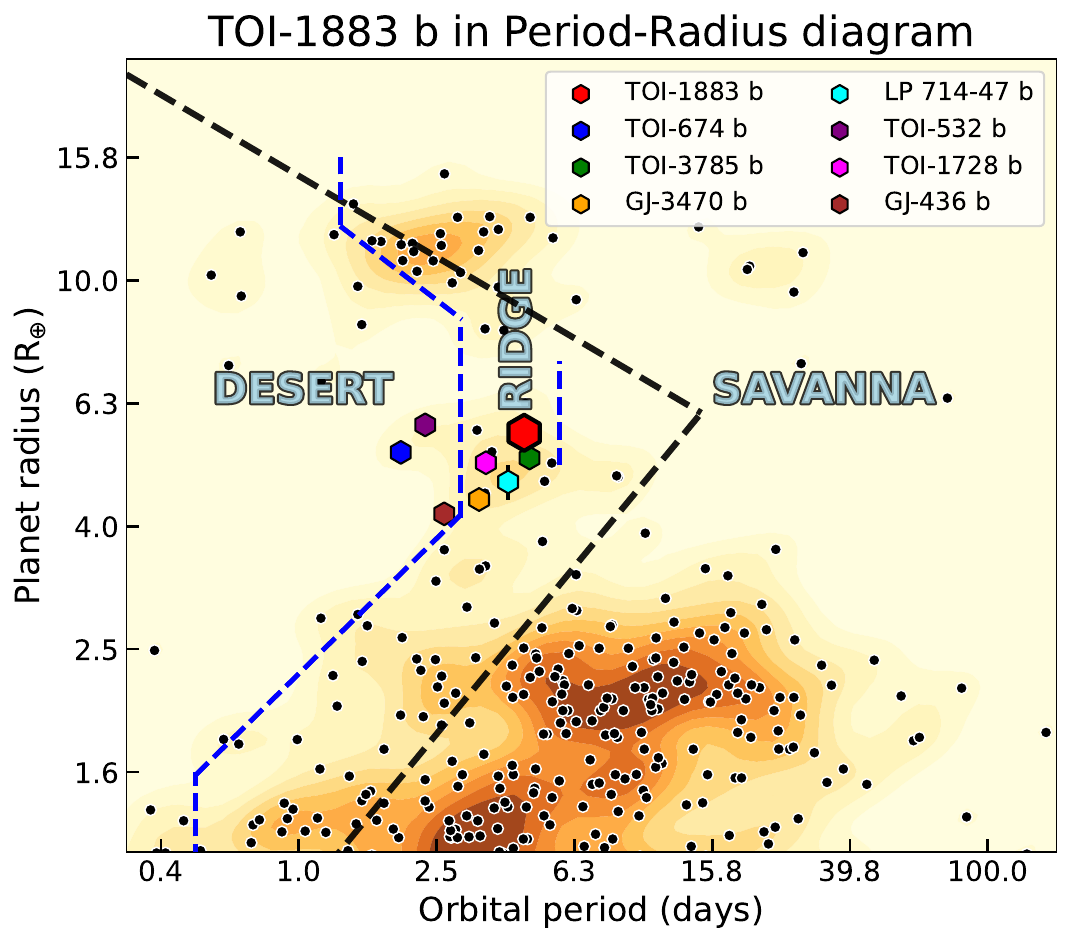}
\end{center}
\caption{
Planets orbiting M-type stars ($T_{\mathrm{eff}} < 4000\,\mathrm{K}$) are plotted in black on the orbital period–planet radius diagram. TOI-1883 b is highlighted in red. 
Planets with mean densities similar to that of TOI-1883\,b are plotted using the same colors as in Figure \ref{fig:MR_1883}.
The black dashed lines indicate the boundaries of the \textit{Neptune desert} defined by \citet{Mazeh16}. The population-based boundaries of the \textit{Neptune desert}, \textit{ridge}, and \textit{savanna} (blue dashed line) are taken from \citet{2024CastroGonzalez}.
We selected planets with well-constrained radii by requiring both the upper and lower uncertainties in $R_{\rm p}$ to be smaller than three times the median uncertainty of the sample.
{Alt text: The planet radius–orbital period diagram is shown, with the boundaries of the \textit{Neptune desert}, \textit{ridge}, and \textit{savanna} overlaid, along with planets orbiting M-type stars.}
} \label{fig:desert}
\end{figure}

\section{Discussion} \label{sec:discussion}
\subsection{TOI-1883 b in Parameter Space}
The mass is derived as $M_{\rm p} = 13.7^{+6.8}_{-6.5} M_{\earth}$, and the radius is $R_{\rm p} = 5.65\pm0.24 R_{\earth}$, from our RV data for a 1-planet circular model and from \citet{2024PelaezTorres}. 
We derive the mean density $\rho_{\rm p}$ = $0.4 ^{+ 0.3 }_{- 0.2}$ g cm$^{-3}$, and the equilibrium temperature assuming albedo values of 0.0 and 0.3 are also calculated to be $T_{\rm eq} = 544\pm28$ and $498\pm26$ $\mathrm{K}$, respectively. 
The parameters of TOI-1883 b derived in this study are summarized in Table \ref{tb:para1883}.

Figure \ref{fig:MR_1883} shows the distribution of planetary mass and radius based on the data retrieved from the NASA Exoplanet Archive (table version accessed on 2025-10-20). For each sample (M-dwarf and FGK-type stars), we selected planets with well-constrained radii by requiring both the upper and lower uncertainties in $R_{\rm p}$ to be smaller than three times the median uncertainty of the sample. 
The red point represents TOI-1883\,b, black points show planets orbiting M dwarfs ($T_\mathrm{eff} < 4000$\,$\mathrm{K}$), and gray points show those around FGK-type stars ($4000$\,$\mathrm{K}$ $<$ $T_\mathrm{eff}$ $<$ $6500$\,$\mathrm{K}$). Planets with mean densities comparable to that of TOI-1883 b and radii in the range of 4–6\,$R_{\oplus}$ are color-coded and plotted.
The dashed lines represent constant mean densities of 0.3, 0.5, 1, and 2~g~cm$^{-3}$ from top to bottom, while the solid lines show the planetary interior structure models from \citet{Zeng19}.
We further enlarge the lower-right region of Figure \ref{fig:MR_1883}, focussing on the parameter space of planetary masses between 10 and 60~$M_{\oplus}$ and radii between 4 and 6~$R_{\oplus}$ to examine planets located near TOI-1883\,b. 
\textcolor{black}{Although the mass and density are constrained only at modest significance, a comparison of the mean densities of these planets suggests that TOI-1883 b lies in the low-density regime of super-Neptunes orbiting M dwarfs.}

Taking these parameters into account, we aim to assess the extent to which a low-density planet can retain its radius against photoevaporative mass loss.
For this purpose, we utilize the publicly available \texttt{photoevolver} code \citep{fernandez2023shared, photoevolver}, which models the coupled thermal and atmospheric mass-loss evolution of close-in exoplanets subjected to high-energy irradiation (e.g., XUV flux) from their host stars. 
Here, the core composition is assumed to be rocky and is adopted from \citet{otegi2020revisited}, the planetary envelope is assumed to consist of H/He, and its structure is defined following \citet{chen2016evolutionary}. Also, an energy-limited mass-loss model with an efficiency of 0.15 is applied.
With this framework, we computed the core mass, envelope mass, core radius, envelope radius, and the envelope-to-core mass fraction of TOI-1883\,b at evolutionary ages of 1, 5, and 10\, Gyr (Table \ref{tab:photoevolver_states}). It is important to note that these results assume the current age of the planet; As a planet evolves, it cools and its envelope contracts. Therefore, for older system ages, reproducing the observed planetary radius requires a smaller core mass, which reduces the planet’s self-gravity and allows the envelope to remain more extended. 

\begin{table}[tb]
\caption{Derived planetary parameters from the \texttt{photoevolver} output\\
for TOI-1883\,b.}
\begin{center}
\begin{tabular}{lccc}
\hline \hline
Parameter & 1 Gyr & 5 Gyr & 10 Gyr \\
\hline
Core mass $M_{\rm core}$ ($M_\oplus$) & $10.2$ & $8.6$ & $7.7$ \\
Envelope mass $M_{\rm env}$ ($M_\oplus$) & $3.5$ & $5.1$ & $6.0$\\
Core radius $R_{\rm core}$ ($R_\oplus$) & $2.0$ & $1.9$& $1.9$ \\
Envelope mass fraction $f_{\rm env}$ & $0.3$ & $0.6$ & $0.8$ \\
Envelope radius $R_{\rm env}$ ($R_\oplus$) & $3.6$ & $3.7$ & $3.8$ \\
\hline
\end{tabular}
\end{center}
\vspace{2mm}
\footnotesize{
\textit{Note.} The envelope mass is computed as 
$M_{\rm env} = M_{\rm tot} - M_{\rm core}$, with $M_{\rm tot} = 13.7\,M_\oplus$. The envelope radius is computed as $R_{\rm env} = R_{\rm tot} - R_{\rm core}$, with $R_{\rm tot} = 5.65\,R_\oplus$. The envelope mass fraction is defined as $f_{\rm env} = M_{\rm env}/M_{\rm core}$. 
}

\label{tab:photoevolver_states}
\end{table}

\subsection{Implications for the \textit{Neptune Desert} and \textit{Ridge}} \label{sec:desert}
Among the planets orbiting M dwarfs, TOI-674\,b \citep{TOI-674, TOI-674_2}, TOI-3785\,b \citep{TOI-3785}, and GJ-3470\,b \citep{GJ3470, GJ3470_2, GJ3470_3, GJ3470_4} have mean densities similar to that of TOI-1883\,b. 
Their reported mean densities are \textcolor{black}{$\rho_{\rm p} \sim 0.9 \pm0.2 ~{\rm g~cm^{-3}}$ for TOI-674\,b, $\rho_{\rm p}\sim0.6 \pm0.2~{\rm g~cm^{-3}}$ for TOI-3785\,b, and $\rho_{\rm p}\sim0.8\pm0.1~{\rm g~cm^{-3}}$ for GJ-3470\,b.} 
Although these planets exhibit comparable bulk densities, their radii are all slightly smaller than those of TOI-1883 b. 
To place TOI-1883\,b and these comparable planets in a broader context, 
we plotted the positions of the planets orbiting M-type stars ($T_\mathrm{eff} < 4000$ $\mathrm{K}$) relative to the previously defined boundaries of the \textit{Neptune desert} \citep{2011SzaboKiss, Mazeh16, Lopez17,2024CastroGonzalez} on the radius--orbital period diagram (Figure~\ref{fig:desert}), following mainly the methodology of \citet{2024CastroGonzalez}.

The colored planets highlighted in Figure \ref{fig:MR_1883} are plotted in the planet radius--orbital period diagram in Figure \ref{fig:desert}. These planets span orbital periods of approximately 2--6 days and are distributed from the \textit{Neptune desert} through the \textit{ridge} region \citep{2011SzaboKiss, Mazeh16, Lopez17,2024CastroGonzalez}.
The previously defined \textit{Neptune desert}, \textit{ridge}, and \textit{savanna} were mainly characterized by planets discovered by the \textit{Kepler} survey \citep{borucki2010kepler} orbiting FGK-type stars \citep{Mazeh16, 2024CastroGonzalez}.
However, as shown in Figure \ref{fig:desert}, the previously defined \textit{Neptune desert} and \textit{ridge} boundaries derived from FGK-type stars also describe the distribution of planets orbiting M dwarfs remarkably well.
This may be because M dwarfs exhibit higher XUV/bolometric flux ratios than solar-type stars. As a result, photoevaporation can efficiently strip planetary atmospheres even at the same orbital period, where the bolometric insolation is lower, shaping the boundary of the \textit{Neptune desert} at similar orbital periods \citep{owen2016habitability,fernandez2024survival,Gaidos24}.
TOI-1883\,b lies precisely on the \textit{ridge}, and the planets with mean densities 
similar to that of TOI-1883\,b are likewise distributed from the \textit{Neptune desert} toward the \textit{ridge}.

In \citet{bourrier2025atreides}, it is argued that, in the orbital period--planetary density plane, low-density Neptunes with $M_{\mathrm{p}} \lesssim 20\,M_{\oplus}$, $R_{\mathrm{p}} \sim 5$--$7\,R_{\oplus}$, and $\rho \sim 0.4$--$1~\mathrm{g\,cm^{-3}}$ (i.e., planetary envelope mass fraction $< 1$) are scarcely found within the \textit{Neptune desert}. This paucity suggests that atmospheric escape begins in the \textit{ridge} and that, as the envelope mass fraction approaches $0.01$ \citep{owen2018photoevaporation}, these planets evolve into more stable sub-Neptunes or even bare rocky cores.
Moreover, the density brink shown in Fig.~A.1 of \citet{bourrier2025atreides} is consistent with the observed trend and can be interpreted in the context of an evolutionary scenario.
In this scenario, low-density Neptunes would have migrated inward via disk-driven migration within $\sim 100$ Myr before disk dispersal, and subsequently experienced photoevaporation under the strong XUV irradiation from their young host star.


Following this framework, TOI-1883\,b, with a bulk density of $\rho \approx 0.4~\,\mathrm{g\,cm^{-3}}$, is consistent with having formed via disk-driven migration. Moreover, if TOI-1883\,b has indeed experienced photoevaporative mass loss, its atmosphere may be enriched in helium or heavy elements. Such enrichment could be tested with future high-precision transmission spectroscopy. Signatures consistent with photo-evaporative atmospheric escape have already been detected for other M-dwarf Neptunes, including neutral hydrogen in GJ~436\,b \citep{hu2015helium, madhusudhan2011high, morley2017forward, lavie2017long} and both neutral hydrogen and the helium triplet in GJ~3470\,b \citep{bourrier2018hubble, palle2020he}.

On the other hand, we find that the orbital circularization timescale is 
$\sim 170$~Myr for a tidal quality factor of $Q_{\rm p} = 10^{4}$ and $\sim 1.7$~Gyr for $Q_{\rm p} = 10^{5}$. Therefore, the possibility that the planet once had a non-zero eccentricity but that it has already been damped cannot be ruled out.

Recent studies suggest that stars hosting planets in or near the \textit{Neptune desert} tend to be more metal-rich than those hosting planets outside this region \citep{beauge2012emerging,dai2021tks, dong2018lamost, doyle2025exploring,vissapragada2025hottest}. Given the measured stellar metallicity of TOI-1883\,b's host star, [Fe/H] $= 0.32 \pm 0.18$~dex, this system is consistent with that tendency. 
In the context of planets residing in the Neptune desert, an enhanced metallicity in the protoplanetary disk increases the optical depth of the gas, resulting in higher opacity that reduces the efficiency of radiative cooling within the envelope, thereby delaying contraction. This effect increases the critical core mass required for runaway gas accretion and may suppress the formation of gas giant planets. \citep{ikoma2000formation, piso2014minimum}.
Applied to TOI-1883\,b, this mechanism allows for a scenario in which the planet experienced photoevaporation during its migration toward the \textit{ridge} before it could grow into a gas giant, leading to its present-day mass–radius properties. This interpretation is also consistent with previous findings that hot Neptunes—particularly single-planet systems—are more readily produced at higher metallicities \citep{petigura2018california,dong2018lamost}.

Therefore, to place more stringent constraints on the formation history of the planet, several observational approaches will be particularly valuable. These include improved radial-velocity measurements for a more precise mass and eccentricity determination, measurements of the projected spin–orbit angle via the Rossiter-McLaughlin effect to constrain the planetary obliquity, and transmission spectroscopy to probe the atmospheric composition and escape. In particular, the particularly high transmission spectroscopy metric (TSM $> 140$) makes TOI-1883\,b an excellent target for future atmospheric characterization with facilities such as \textit{JWST} \citep{Hord24} and \textit{Ariel} 
\citep{Edwards22}.

\section{Conclusion} \label{sec:conclusion}
We report on follow-up observations of a planetary system with a transiting short-period super-Neptune around the mid-M dwarf TOI-1883. The transiting planet TOI-1883~b is a confirmed super-Neptune with an orbital period of $P=4.508$ days, making it a suitable target for atmospheric observation.
We measured the planetary mass and stellar properties with the IRD mounted on the Subaru telescope and obtained the stellar and planetary properties from additional transit observations by the TESS and ground-based multicolor photometry with MuSCAT2 and MuSCAT3. 
The planetary mass of TOI-1883~b is determined to be $M_{{\rm p}} = 13.7 ^{+ 6.8 }_{- 6.5 } M_{\earth}$ from the RV data, and the mean density is derived to be $\rho_{\rm p}$ = $0.4 ^{+ 0.3 }_{- 0.2}$ g cm$^{-3}$.
Among the super-Neptunes discovered around M-type stars, TOI-1883\,b is the lowest-density planet known, and is therefore likely to be an extremely inflated, ``puffy'' planet. 
It also resides on the \textit{ridge}, as statistically defined from FGK-type star samples. Given its location on the \textit{ridge} and its low bulk density, it is suggested that the planet may have migrated inward via disk-driven migration and subsequently experienced photoevaporative mass loss driven by stellar XUV irradiation. 
Furthermore, the super-solar metallicity of the host star may have inhibited runaway gas accretion, preventing TOI-1883\,b from forming as a gas giant while it migrated inward.
Future atmospheric characterization and high-precision radial velocity measurements will enable stronger constraints on the formation and evolutionary history of this system. The density and migration scenario inferred in this study provides important insight into the origins and evolution of short-period Neptune-sized planets residing in the \textit{ridge} region.

\begin{ack}
This research is mainly based on data collected at the Subaru Telescope, which is operated by the National Astronomical Observatory of Japan. We are honored and grateful for the opportunity of observing the Universe from Maunakea, which has the cultural, historical and natural significance in Hawaii.
This paper is based on observations with the MuSCAT2 instrument, developed by Astrobiology Center, at Telescopio Carlos S\'{a}nchez operated on the island of Tenerife by the IAC in the Spanish Observatorio del Teide.
This paper is also based on observations with the MuSCAT3 instrument, developed by Astrobiology Center and under financial supports by JSPS KAKENHI (JP18H05439) and JST PRESTO (JPMJPR1775), at Faulkes Telescope North on Maui, HI, operated by the Las Cumbres Observatory.
This work makes use of observations from the Las Cumbres Observatory global telescope network and supported by the JSPS KAKENHI Grant Numbers This study was partly supported by the JSPS KAKENHI Grant Numbers JP19KK0082, JP20K14521, JP21K13955, JP21K13987, JP22H05150, JP23H00133, JP23H01224, JP23H01227, JP23K17709, JP23K25920, JP23K25923, JP24H00017, JP24H00242, JP24H00248, JP24K00689, JP24K07108, JP24K17082, JP24K17083, JP25H00005, JP25K01061, JP25K17450, JSPS Bilateral Program Number JPJSBP120249910, JSPS Grant-in-Aid for JSPS Fellows Grant Number JP24KJ0241, JP25KJ0091, JP25KJ1036, JP25KJ1040, JST SPRING Grant Number JPMJSP2108.
F.M. acknowledges the financial support from the Agencia Estatal de Investigaci\'{o}n del Ministerio de Ciencia, Innovaci\'{o}n y Universidades (MCIU/AEI) through grant PID2023-152906NA-I00.
\end{ack}

\appendix 

\section{ASAS-SN light curves and their periodogram} 
\label{apn:activity_analysis}
\begin{figure*}[htb]
\begin{center}
\includegraphics[width=10cm]{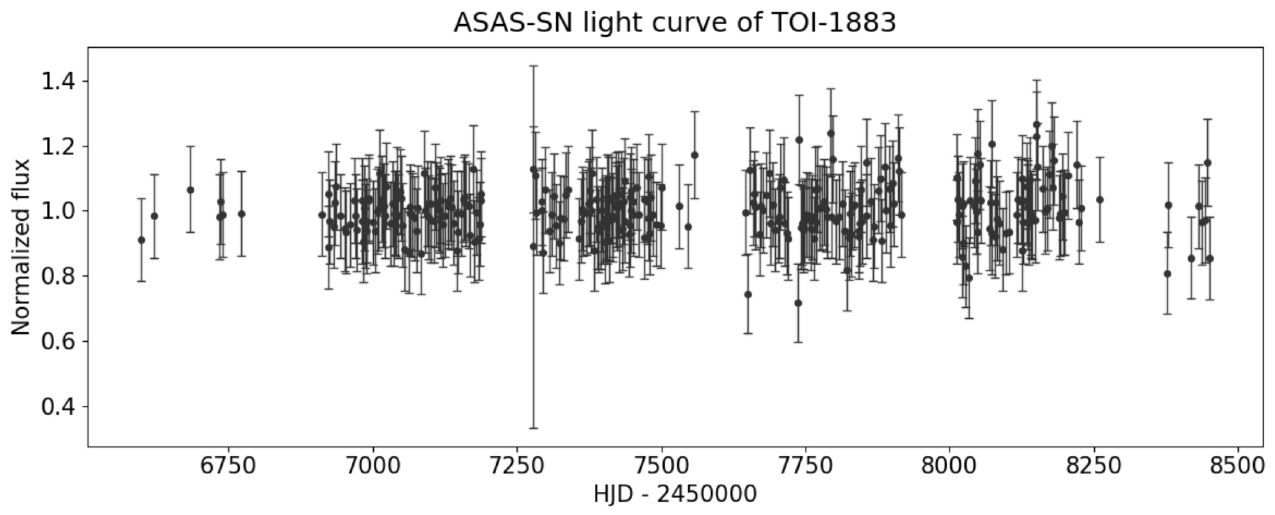}
\end{center}
\caption{The light curve of TOI-1883 observed by ASAS-SN in $V$-band (Section \ref{sec:activity}).
{Alt text: The light curve of TOI-1883 observed by ASAS-SN in $V$-band.}
}
\label{fig:asas-sn_lc}
\end{figure*}

\begin{figure*}[htb]
\begin{center}
\includegraphics[width=10cm]{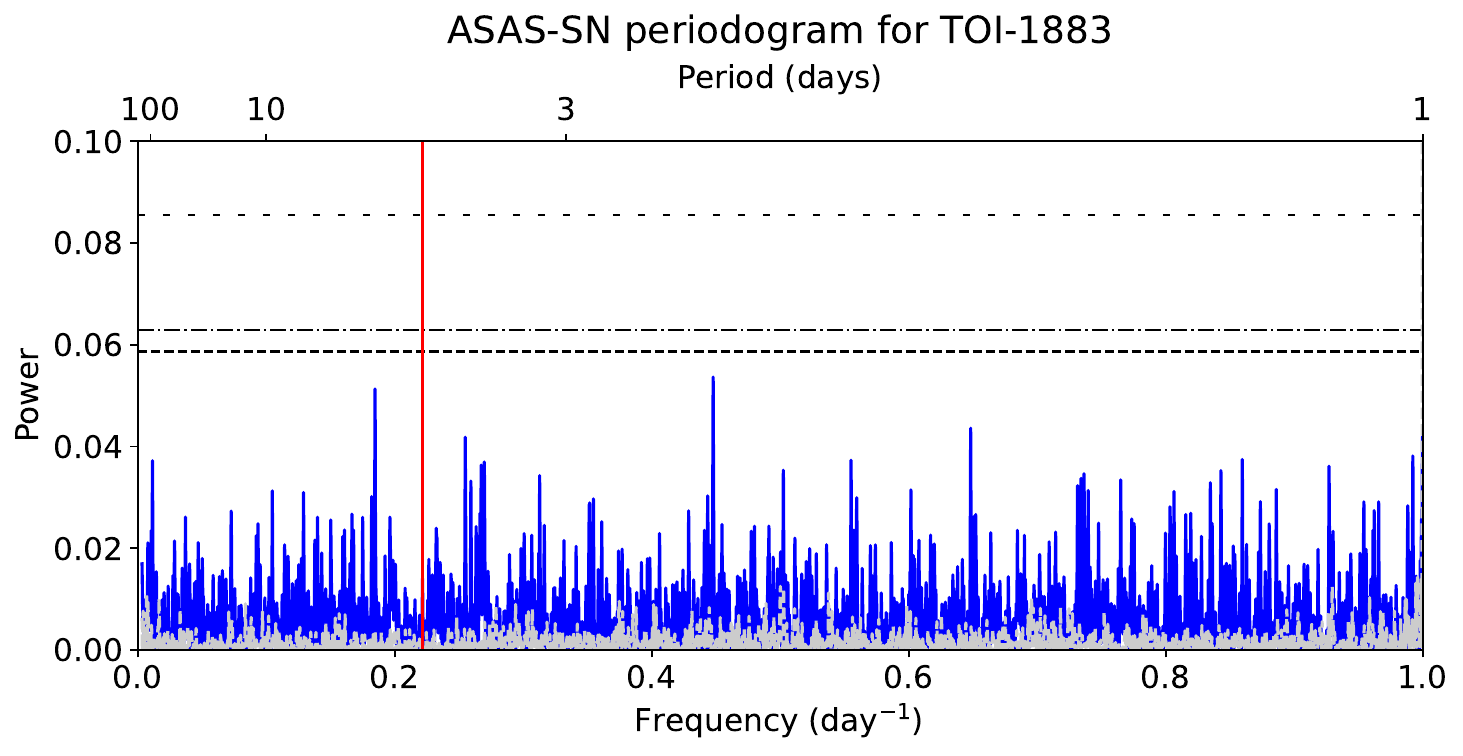}
\end{center}
\caption{
Periodograms for the ASAS-SN light curves in $V$-bands (blue), and their window functions (gray), with the GLS for TOI-1883 (Section \ref{sec:activity}).
The vertical line represents the orbital period (=4.506 days) of the planet (red), and the horizontal lines represent the FAP of 0.1, 5.0, and 10.0 \%, respectively (black).
{Alt text: The period analysis of the ASAS-SN light curve is presented.}
}
\label{fig:asas-sn_period}
\end{figure*}

Figure \ref{fig:asas-sn_lc} shows the ASAS-SN light curves in $V$-band. There is no significant photometric variability ascribed to stellar activity. Figure \ref{fig:asas-sn_period} shows the GLS periodogram for the ASAS-SN light curve (Section \ref{sec:activity}).

\section{Corner plot for the 1-planet RV model}
\label{apn:1-p RV corner}
\begin{figure*}[htb]
\begin{center}
\includegraphics[width=10.5cm]{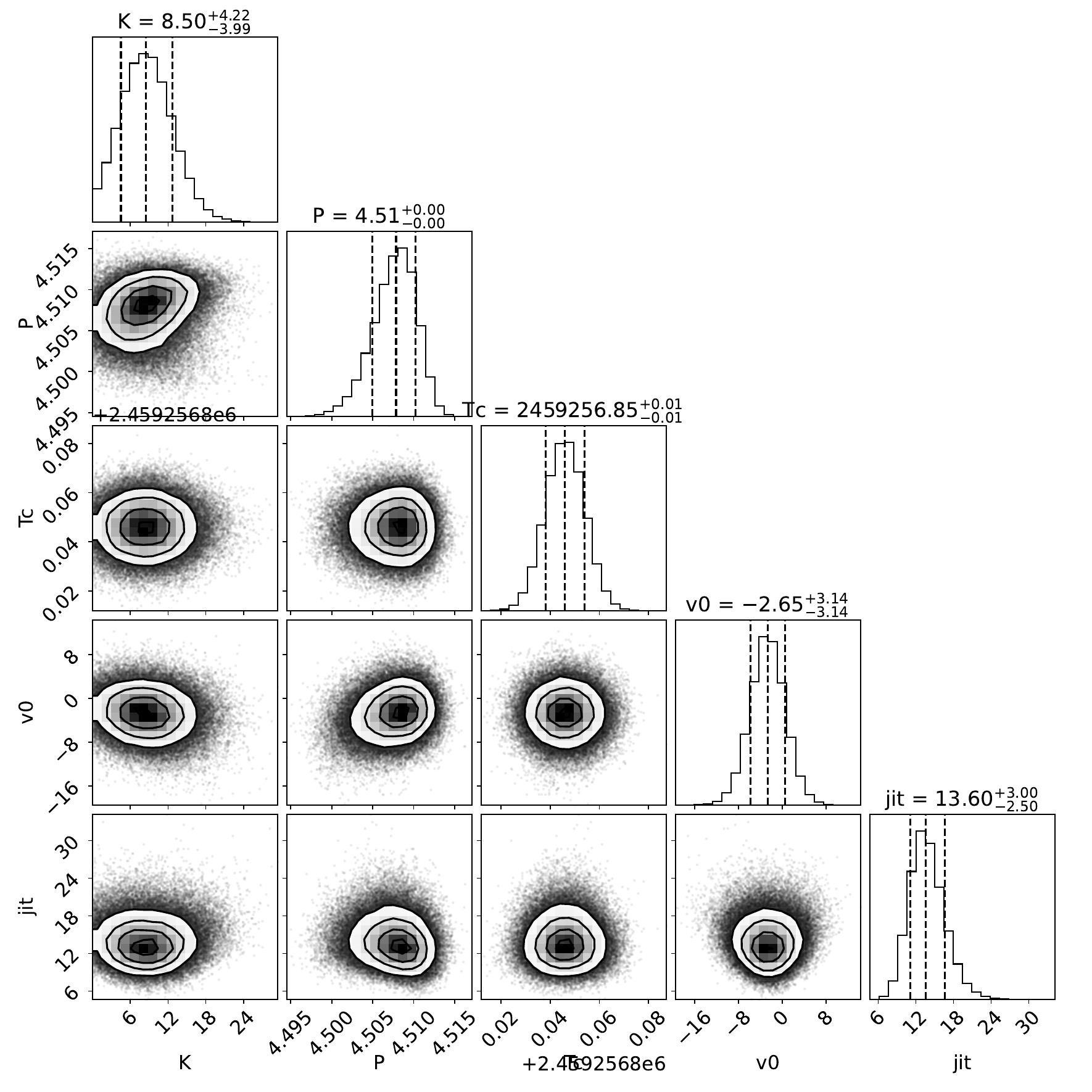}
\end{center}
\caption{This is a corner plot of the posterior distribution for each
parameter of the 1-planet circular orbit model. This result was gen-
erated from an MCMC with 50000 iterations and 10000 discards.
{Alt text: This is a corner plot of the posterior distribution for each
parameter of the 1-planet circular orbit model.}
}  
\label{fig:corner}
\end{figure*}

Figure \ref{fig:corner} shows the corner plots of the posterior probability distributions for the 1-planet circular RV model.

\bibliography{ird}

\bibliographystyle{aasjournal}
\end{document}